\documentclass[twocolumn]{aastex631}

\usepackage {soul}
\usepackage{tocbibind}
\usepackage[toc,page]{appendix}
\usepackage{amsmath}

\newcommand{\msun}{\,\rm M_\odot}
\definecolor{ejcol}{rgb}{0,0,0}

\shorttitle{Star formation variability}
\shortauthors{Shin et al.}
\graphicspath{{./}{figures/}}

\begin{document}

\title{Star formation variability as a probe for the baryon cycle within galaxies}

\author[0000-0002-4639-5285]{Eun-jin Shin}
\affiliation{Center for Theoretical Physics, Department of Physics and Astronomy, Seoul National University, Seoul 08826, Korea}

\correspondingauthor{Eun-jin Shin}
\email{shinej816@snu.ac.kr}

\author[0000-0002-8224-4505]{Sandro Tacchella}
\affiliation{Kavli Institute for Cosmology, University of Cambridge, Madingley Road, Cambridge, CB3 0HA, UK}
\affiliation{Cavendish Laboratory, University of Cambridge, 19 JJ Thomson Avenue, Cambridge, CB3 0HE, UK}

\author[0000-0003-4464-1160]{Ji-hoon Kim}
\affiliation{Center for Theoretical Physics, Department of Physics and Astronomy, Seoul National University, Seoul 08826, Korea}
\affiliation{Seoul National University Astronomy Research Center, Seoul 08826, Korea}

\author[0000-0001-9298-3523]{Kartheik G. Iyer}
\affiliation{Dunlap Institute for Astronomy and Astrophysics, 50 St. George Street, Toronto, Ontario M5S 3H4, Canada}
\affiliation{Columbia Astrophysics Laboratory, Columbia University, 550 West 120th Street, New York, NY 10027, USA}

\author[0000-0002-6648-7136]{Vadim A. Semenov}
\affiliation{Center for Astrophysics $|$ Harvard \& Smithsonian, 60 Garden St, Cambridge, MA 02138, USA}



\begin{abstract}
We investigate the connection of the regulation of star formation and the cycling of baryons within and in and out of galaxies. We use idealized numerical simulations of Milky Way-mass galaxies, in which we systemically vary the galaxy morphology (bulge-to-total mass ratio) and stellar feedback strength (total eight setups with 80 simulations). 
By following individual gas parcels through the disk, spiral arms, and massive star-forming clumps, we quantify how gas moves and oscillates through the different phases of the interstellar medium (ISM) and forms stars. 
We show that the residence time of gas in the dense ISM phase ($\tau_{\rm SF}$), the nature of spiral arms (strength, number), and the clump properties (number, mass function, and young star fraction) depend on both the galaxy morphology and stellar feedback.
Based on these results, we quantify signatures of the baryon cycle within galaxies using the temporal and spatial power spectrum density (PSD) of the star formation history (SFH).
Stronger stellar feedback leads to more bursty star formation while the correlation timescale of the SFH is longer, because stronger feedback dissolves the dense, star-forming ISM phase, leading to a more homogeneous ISM and a decrease in $\tau_{\rm SF}$. 
The bulge strength has a similar effect: the deep gravitational potential in a bulge-dominant galaxy imposes a strong shear force that effectively breaks apart gas clumps in the ISM; this subsequently inhibits the fragmentation of cool gas and therefore the star formation in the disk, leading to a decrease in the spatial power on scales of $\sim$ 1 kpc.
We conclude that measurements of the temporal and spatial PSD of the SFH can provide constraints on the baryon cycle and the star formation process.

\end{abstract}
 


\section{Introduction} \label{sec:intro}
Star-forming galaxies are dynamic ecosystems in which gas cycles in and out of the disk, governed by a wide range of physical processes that act from stellar to cosmological scale: the growth of the large-scale structure, the cooling and heating of interstellar medium (ISM), and the formation of stars and central black holes and their associated feedback processes.
Regulator models provide a holistic view of galaxy evolution using a simple fundamental continuity equation for the content of gas and stellar mass.
Based on the balance between gas inflows, outflows, star formation, and recycling, this model reproduces global scaling relations of galaxies for the overall baryon content of galaxies \citep{Bouche2010, Dave2012, Dekel2014}, the fundamental metallicity relation \citep{Lilly2013}, and evolution with oscillations along the star-forming main sequence \citep{Tacchella2016}.
Despite the success of this model, we still lack detailed understanding of how star formation and feedback operate from the stellar to the cosmic scale. 

Numerical simulations are a powerful tool for exploring the details of how star formation and feedback processes interact within galaxies and their surrounding gas. 
Cosmological hydrodynamic simulations demonstrate how baryons cycle throughout the galaxies' cosmic evolution: galactic gas accretion \citep{Keres+2005,Faucher-Giguere+2011,Putman2012,Fraternali2017}, galactic outflows driven by stars and black holes \citep{Oppenheimer+Dave2008, Veilleux+2005, Heckman+Thompson2017,Forster-Schreiber+2014, Angles-Alcazar+2014,Forster-Schreiber+2019} and re-accretion back onto galaxies \citep{Oppenheimer+2010, Christensen+2016, Angles-Alcazar+2017} all play a fundamental role in the growth of galaxies.
For those cosmological simulations, however, feedback models are implemented using subgrid approaches that include free parameters, such as hydro-dynamically decoupled wind particles \citep{Hopkins+2012,Fielding+2017,Li+Tonnesen2020}, effective ISM equation of state \citep{Yepes+1997,Springel+Hernquist2003,Braun+Schmidt2012}, or the temporary shutdown of cooling \citep{DallaVecchia+Schaye2012,Stinson+2013}, to name a few.

Non-cosmological, idealized simulations offer the possibility to inspect the interplay of gas and stellar feedback in galaxies in a more controlled environment with a significant higher resolution (by a factor $\sim10-100$).
The typical astrophysical conditions, including shocks, shear, and multi-phase ISM, are resolved and fewer subgrid models are required \citep{McKee+Ostriker2007, Kim+Ostriker2015}. 
The molecular cloud physics \citep{Padoan+Nordlund2002,Kim+2003, Bate+Bonnell2005,Hennebelle+Chabrier2008,Guszejnov+2021,Dobbs+2022}, photo-ionization and photo-electric radiation pressure \citep{Kim+2013,Hu+2017,Emerick+2018} can also be modeled, which are important since they control the star formation and the ISM turbulence. 
Moreover, the dynamical and morphological structure, including spiral arms and bars, naturally arise and can enhance and quench the star formation \citep{Martig+2009,Elmegreen2011,Gensior2020,Dobbs+2022}.

To constrain the physical properties of galaxy evolution, it is necessary to comprehend the star formation history (SFH) of galaxies, which is the temporal and spatial record of the various physical processes on star formation and galaxy growth.
To quantify the variability of SFHs in galaxies, \cite{Caplar+Tacchella2019} proposed the use of the power spectrum density (PSD) of the SFH and presented a first measurement of the PSD in local, Milky Way-like galaxies. 
\cite{Wang+Lilly2020} investigated the observational constraints of SFH PSD based on SDSS-IV MaNGA data.
\cite{Tacchella2020} studied how distinct physical processes give rise to different SFH PSDs by extending the regulator model by \cite{Lilly2013} to the giant molecular cloud (GMC) scale. Together with \citet{Semenov2017,Semenov2018}, a picture arises where star formation variability directly relates to the characteristic timescale of the baryon cycling driven by stellar feedback and the creation and destruction of GMCs.
Additionally, \cite{Iyer2020} examined the PSDs of SFHs extracted from cosmological simulations, zoom-in simulations, and semi-analytical models. 
They reported vast discrepancies in PSDs between simulations on short timescales ($\sim100$ Myr), which indicates that the star formation rate (SFR) fluctuations on short timescales are sensitive to the implemented sub-grid physics and also highlights that we do not well understand the small-scale baryon processes at present.

Motivated by \cite{Iyer2020}, we focus in this work on how the physical processes drive star formation variability, inspecting the detailed gas flows within galaxies using a suite of idealized simulations of Milky Way-mass galaxies.
We take advantage of the particle-based nature of our smoothed particle hydrodynamics (SPH) simulations to track and study gas particles' movement in and out of the disk.
Varying the stellar feedback energy and morphology of the galaxies, we correlate the spatial and temporal PSD features with their underlying physical mechanisms.

The remainder of the paper is structured as follows.
In Section~\ref{sec:simulations}, we describe the simulations that we have carried out for this study to explore the connection between the baryon cycle and stellar physics across different galactic environments. 
In Section~\ref{sec:spatial}, we compare our simulations to key observable properties (such as spiral arms and star-forming clumps) and analyze the spatial distribution of gas and stars in different setups. 
In Section~\ref{sec:temporal}, we analyze how baryons cycle by tracking individual gas parcels through the galaxies and quantify the movement of gas in the temperature-density phase diagram.
Then, we measure the temporal PSD of the SFH and study the connection between the gas dynamics and star formation in different simulated setups.
In Section~\ref{sec:discussion}, we discuss the observable properties related to our work, explore the implications of this study, and additionally state the caveats of our simulations.
Finally, we present a summary of our conclusions in Section~\ref{sec:conclusions}.

\section{Simulations}
\label{sec:simulations}
This section describes the simulations we have carried out for this study.
Section~\ref{subsec:init_conditions} details the initial conditions, and Section~\ref{subsec:simulation_physics} summarizes our simulation setup and baryon physics related to the thermodynamics of the gas and stellar evolution.
We present a suite of 80 simulations with a range of stellar feedback implementations (variation in supernovae energy input) and different initial conditions (variation of the bulge strength relative to the disk) in Section~\ref{subsec:simulation_suite}.

\subsection{Initial Conditions}
\label{subsec:init_conditions}
We use an initial condition provided by the {\it AGORA} Project \citep{Kim+2016}, which contains an isolated disk with the characteristics of a Milky Way-mass galaxy at $z\sim1$ with $M_{\rm 200, crit} = 1.074 \times 10^{12}\msun$.
Specifically, an exponential disk and a stellar bulge following the Hernquist profile \citep{Hernquist1990} are embedded into a dark matter halo that follows the Navarro--Frenk--White profile \citep[NFW;][]{Navarro1997}.
The galaxy consists of a dark matter halo that has a mass of 1.25$\times10^{12}\msun$, a stellar component with a mass of 3.87$\times10^{10}\msun$ ($M_{\rm disk}+M_{\rm bulge}$), and a gas disk with a mass of 8.59$\times10^9\msun$.
We assume a disk scale radius of $r_{\rm d}$ = 3.43 kpc, a disk scale height of $z_{\rm d}=$ 0.1$\,r_{\rm d}$, and a stellar bulge with a scale radius of 0.4 kpc.

To explore the impact of galaxy morphology on the baryon and star formation properties, we vary bulge strengths, while fixing the total stellar mass ($M_{\rm disk}+M_{\rm bulge}$) and the scale lengths of the disk and bulge.
We set the {\it AGORA} initial condition ($\mathrm{B/T}$ = 1/9) as our fiducial setup and test $\mathrm{B/T}$ ratios of 1/30 ({\tt B0.3}), 1/3 ({\tt B3}) and 2/3 ({\tt B6}).
Fixing gas and dark matter components in {\it AGORA} initial condition, we regenerate the position and velocity distributions of disk and bulge components in a dynamical equilibrium using {\sc Dice} code \citep{Perret2016} under the same mass resolution of stellar particles.
We employ 10$^5$ particles for both the dark matter and the gas components, and use 1.13$\times10^5$ particles for the collisionless stellar component.
Each type of particle has a mass of $m_{\rm DM}= $ 1.26$\times10^7\msun$, $m_{\star}=$ 3.44$\times10^5\msun$ and $m_{\rm gas}=$ 5.93$\times10^3\msun$. 
The galaxy also includes a hot gaseous halo---4$\times10^3$ gas particles following the NFW profile, which is necessary for constructing the hot phase in the galaxy in particle-based codes \citep{Shin2021}. 
We set an initial metallicity of $Z_{\rm disk} = 0.02041$ in the disk and $Z_{\rm halo} = 10^{-6}\,Z_{\rm disk}$ for the gas halo.

\subsection{Simulation Setup}
\label{subsec:simulation_physics}

The simulations described in this paper are run with the {\sc Gizmo} code \citep{Hopkins2015} and we analyze the simulated data with the {\tt yt}-toolkit \citep{Turk2011}.
The hydrodynamics is solved with Lagrangian framework using the Pressure-Smoothed Particle Hydrodynamics (Pressure-SPH) scheme \citep{Hopkins2013}. 
We implement the cubic spline kernel \citep{Hernquist1989} for the softening of the gravitational force with $N_{\rm ngb}=32$ for the desired number of neighboring particles.
We set the Plummer equivalent gravitational softening length $\epsilon_{\rm grav}$ to 80 pc and the minimum hydrodynamic smoothing length to $0.2 \epsilon_{\rm grav}$.

\subsubsection{Cooling, heating, pressure floor}

Radiative cooling is modeled using the Grackle-chemistry and cooling library \citep{Smith2017}, which solves non-equilibrium primordial chemistry and cooling for a given metallicity of the gas.
The library also includes tabulated rates of metal cooling calculated with the photoionization code {\sc Cloudy} \citep{Ferland2013} and photoheating and photoionization from the ultra-violet background (UVB) radiation. We adopt the UVB value at $z =$ 0 from \citet{HaardtMadau12}.
We apply a non-thermal Jeans pressure floor that forces the local Jeans length to be resolved to avoid artificial numerical fragmentation \citep{Truelove1997, Kim+2016}:
\begin{equation}
    P_{\rm Jeans} = \frac{G}{\gamma\pi}N^{2}_{\rm Jeans}\rho^{2}_{\rm gas}\Delta x^{2},
\end{equation}
where the adiabatic index $\gamma$ = 5/3, the Jeans number $N_{\rm Jeans}$ = 6.3\footnote{We adopt $N_{\rm Jeans}$ = 6.3 rather than the usual 4, based on the different definition of $h_{\rm sml}$ in {\sc Gizmo} code.}, $G$ is the gravitational constant, $\rho_{\rm gas}$ is the gas density, and $\Delta x$ is the radius of the effective volume of a cell, given by $(4\pi/(3 N_{\rm ngb}))^{1/3}h_{\rm sml}$, where $h_{\rm sml}$ is the smoothing length.

\subsubsection{Stellar physics}

Gas parcels that are denser than a threshold, $\rho_{\rm SF, thres}=1.67\times10^{-23}{\rm g\,cm}^{-3}$ ($n_{\rm H}=10\,{\rm cm}^{-3}$), form stars at a rate following the local Schmidt-like relation
\begin{equation}
    \frac{{d\rho_*}}{dt} = \frac{\epsilon_*\rho_{\rm gas}}{t_{\rm ff}},
\label{eq:SF}
\end{equation}
where $\rho_*$ is the stellar density, $t_{\rm ff} = (3\pi/(32\, G\rho_{\rm gas}))^{1/2}$ is the local free-fall time, and $\epsilon_* = 0.01$ is the star formation efficiency per free-fall time.

Star particles inject thermal energy, mass, and metals into their surrounding ISM 5 Myr after their formation, in an attempt to describe Type II SN explosions.   
Following \citet{Chabrier2003} initial mass function (IMF), we assume that for stars with a mass range of 8 -- 40 $\msun$, a single supernovae event occurs per every 91 $\msun$ of stellar mass formed releasing 2.63 $\msun$ of metals and 14.8 $\msun$ of gas (including metals).
For the {\tt Fiducial} run, we inject thermal energy of 10$^{51}$ ergs per SN event, and we boost the thermal energy by a factor of 2 to 10 to investigate how the feedback strength affects star formation and the baryon cycle within galaxies (see Table \ref{tab:simulations} and Section \ref{subsec:simulation_suite} for more details).
Although such a thermal feedback model is known to suffer from artificially enhanced cooling loses, we still find a significant effect of the feedback boost factor on the SFR magnitude and burstiness. Thus, this thermal feedback model is sufficient for exploring the effects of feedback strength qualitatively. Feedback implementations that are less severely affected by the over-cooling would produce even stronger effects. 

\begin{deluxetable}{ccccc}
\centering
\tabletypesize{\footnotesize}
\tablecolumns{3}
\tablewidth{0pt}
\tablecaption{List of simulations and key parameters.}
\tablehead{\colhead{Setup name}&\colhead{Stellar feedback}&\colhead{Bulge mass}&\colhead{disk mass}\\
\colhead{}&\colhead{[10$^{51}$erg/SN]}&\colhead{$[10^9\msun]$ }&\colhead{$[10^9\msun]$}}
\startdata
{\tt Fiducial}& 1 &4.3& 34.4\\
{\tt FB2}  & 2& 4.3& 34.4 \\
{\tt FB3} & 3 &4.3& 34.4\\
{\tt FB4}& 4 &4.3& 34.4\\
{\tt FB10}&10& 4.3& 34.4\\
{\tt B0.3}&1 &1.29& 37.4\\
{\tt B3}&1 &12.9&25.8\\
{\tt B6}&1 &25.8&12.9\\
\enddata
\tablecomments{List of simulations with different bulge strengths and thermal stellar feedback energy. Each simulation setup is run 10 times with varying initial random seeds, leading to a total of 80 simulation runs.
\label{tab:simulations}
}
\end{deluxetable}

\begin{figure*}
\centering
\includegraphics[width=16cm]{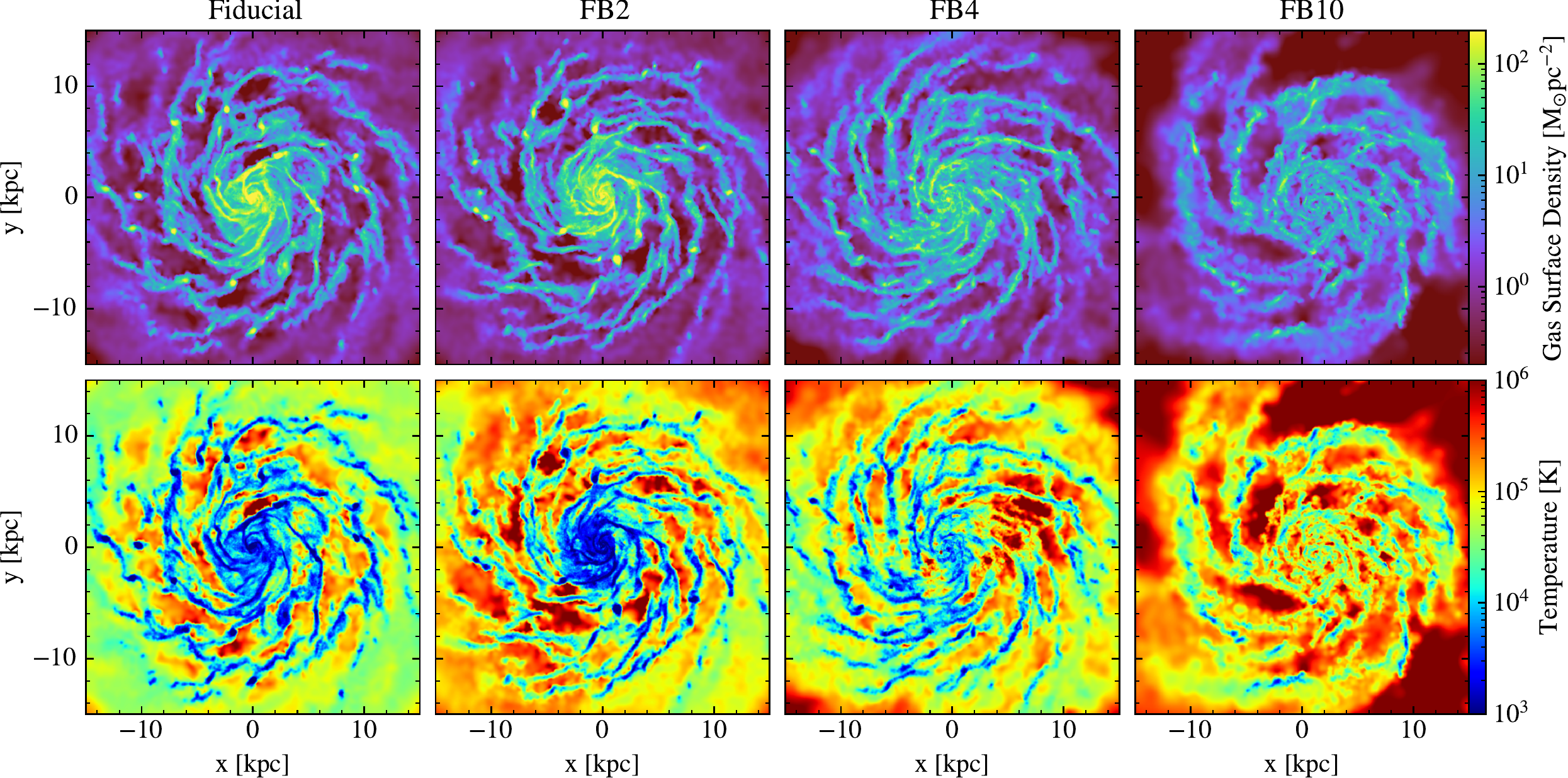}
\caption{Maps of gas surface density ({\it top} panels) and density-weighted gas temperature ({\it bottom} panels) of our isolated Milky Way-mass galaxy simulations with different stellar feedback energies. All simulations have the same initial conditions and all panels show the snapshot at $t=$ 500 Myr. The stellar feedback strength increases from left to right: $10^{51}$ erg/SN ({\tt Fiducial}), $2\times10^{51}$ erg/SN ({\tt FB2}), $4\times10^{51}$ erg/SN ({\tt FB4}), and $10^{52}$ erg/SN ({\tt FB10}). See Table~\ref{tab:simulations} and Section~\ref{subsec:simulation_suite} for the details of the runs. 
The simulations with higher stellar feedback energy show more diffuse and hotter spiral arms, and fewer gas clumps.}
\label{fig:maps_fb_vary}
\end{figure*}

\begin{figure*}
\centering
\includegraphics[width=16cm]{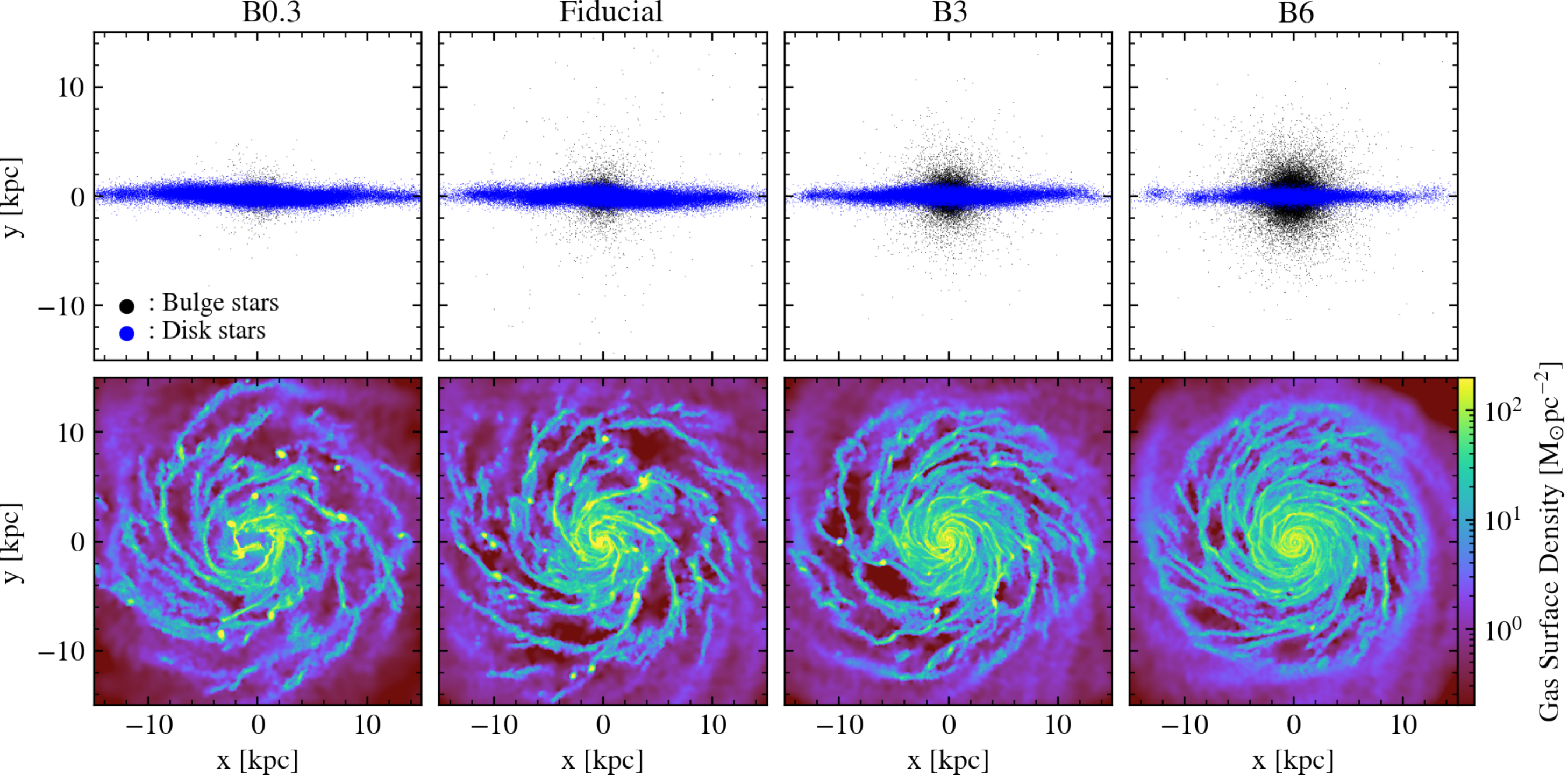}
\caption{Edge-on projection of the stellar particles in the {\it top} panels and the face-on projection of the gas surface density in the {\it bottom} panels. 
Increasing bulge strength is shown from left to right: $\mathrm{B/T}=0.03$ ({\tt B0.3}), $\mathrm{B/T}=0.11$ ({\tt Fiducial}), $\mathrm{B/T}=0.33$ ({\tt B3}), and $\mathrm{B/T}=0.66$ ({\tt B6}), as specified in  Table~\ref{tab:simulations} and Section~\ref{subsec:simulation_suite}.
In the {\it top} panel, the {\it blue} and {\it black} points represent stellar particles in the disk and the bulge, respectively.
All panels show the snapshot at $t=$ 500 Myr.
The shape of the spiral arms and the number of gas clumps in the disk depend on the bulge mass: the disk-dominated galaxy ({\it left} panels) has sharp and high-density spiral arms and more gas clumps than the bulge-dominated galaxy ({\it right} panels).}
\label{fig:maps_bulge_vary}
\end{figure*}

\begin{figure*}
\centering
\includegraphics[width=11cm]{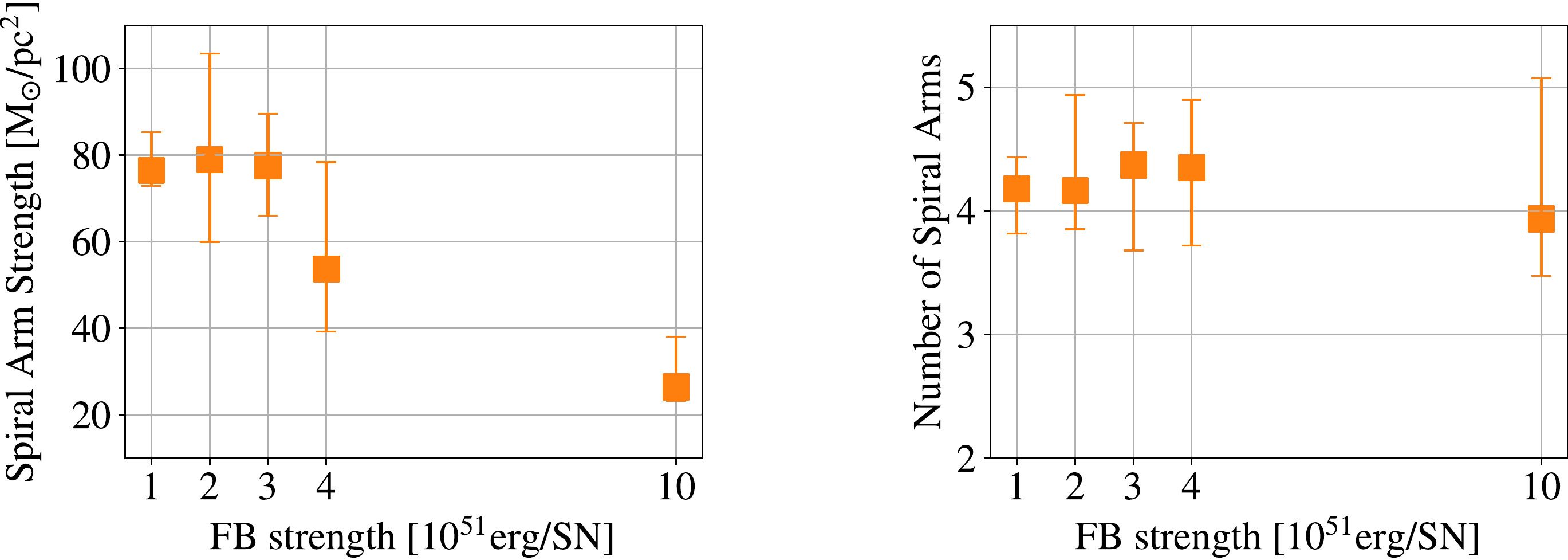}
\includegraphics[width=11cm]{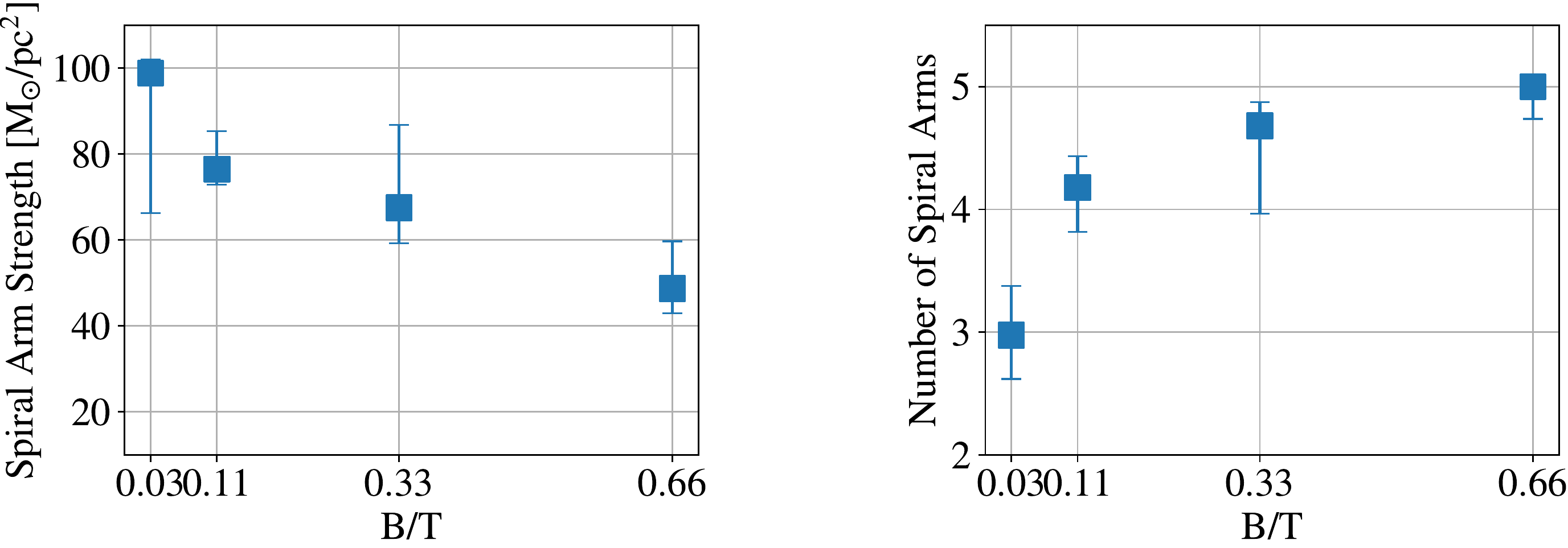}
\caption{Dependence of the spiral arm shape of the galactic disk on stellar feedback strength and bulge strength. 
We plot the spiral arm strength ({\it left} panels) and the number of spiral arms weighted by the spiral arm amplitude at $t=$ 200 Myr ({\it right} panels) and $r=$ 4 kpc as a function of stellar feedback strength ({\it top} panels) and $\mathrm{B/T}$ ratio ({\it bottom} panels). Stronger stellar feedback significantly reduces the spiral arm amplitude, but the number of spiral arms stays approximately the same. Increasing $\mathrm{B/T}$ ratio both decreases the spiral arm amplitude and increases the number of spiral arms. See Section~\ref{subsec:spiral} for details.}
\label{fig:spiral-arm}
\end{figure*}

\begin{figure*}
\centering
\includegraphics[width=17cm]{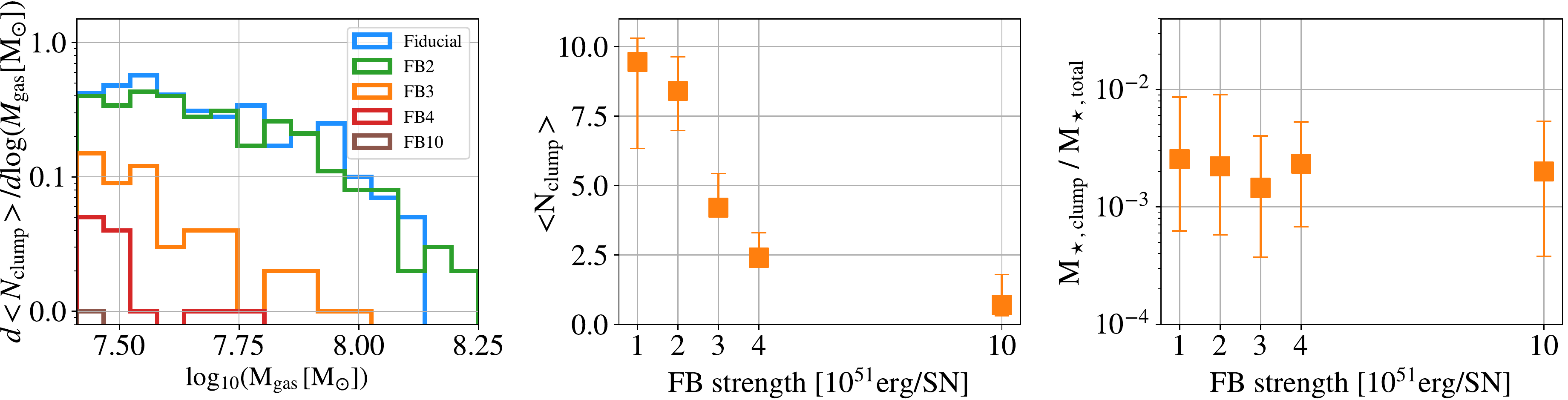}
\includegraphics[width=17cm]{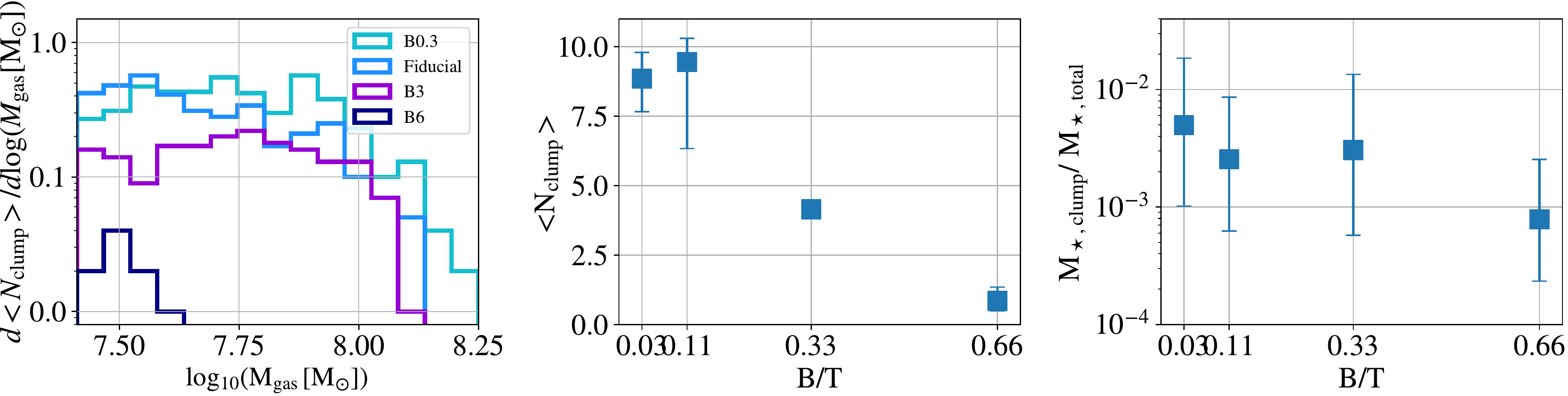}
\caption{
Mass distribution functions of gas clumps in the galaxy and dependence of the number of gas clumps and fraction of young stars in clumps on stellar feedback and bulge strength. 
We plot the time-averaged mass distribution functions of gas clumps in the galaxy ({\it left} panels), time-averaged number of clumps ({\it middle} panels) and the ratio of young star mass ($<$10 Myr from the clump identification; {\it right} panels) in the clump and young star mass in the galaxy as a function of stellar feedback strength ({\it top} panels) and $\mathrm{B/T}$ ratio ({\it bottom} panels). 
The gas clump is identified by {\sc Hop} halo finder algorithm for t = 200 -- 1000 Myr snapshots with 100 Myr timestep.
Increasing feedback strength and higher $\mathrm{B/T}$ ratio both suppress the formation of gas clumps. 
Higher $\mathrm{B/T}$ ratio also reduces the fraction of star formation in clumps, while changing feedback strength does not affect this fraction.
See Section~\ref{subsec:clump} for details.
}
\label{fig:clump}
\end{figure*}

\subsection{Suite of simulations}
\label{subsec:simulation_suite}

We summarize our suite of simulations in Table~\ref{tab:simulations}.
We test eight different kinds of setups with five different stellar feedback parameters (10$^{51}$ ({\tt Fiducial}), 2$\times10^{51}$, 3$\times10^{51}$, 4$\times10^{51}$, and 10$^{52}$ erg/SN) and four different $\mathrm{B/T}$ ratio (1/30, 1/9 ({\tt Fiducial}), 1/3 and 2/3). 
Varying the stellar feedback energy and the $\mathrm{B/T}$ ratio, which directly affect the thermodynamical structure of the ISM and the gravitational potential in the galaxy, respectively, we study how the baryon cycle depends on those effects.
Moreover, both parameters are expected to impact star formation efficiency, so these variations also allows us to approximate the dependence of star formation efficiency on these parameters in the simulations.
We also implement ten different initial random seeds for each setup (i.e. we run a total of 80 simulations) in order to assess the impact of the stochasticity introduced by the numerics \citep{Keller+2019}.
The time interval of snapshots is $\Delta t=$ 10 Myr, while the entire simulation time is 1 Gyr. 
An exception where we only focus on a single run per setup is for the analysis in Section~\ref{subsec:tracking}; we extract 1000 snapshots with a time interval of $\Delta t=$ 1 Myr.

Figure \ref{fig:maps_fb_vary} and \ref{fig:maps_bulge_vary} present two axes of our study: the impact of stellar feedback energy and morphology ($\mathrm{B/T}$ ratio) on the baryonic system in the galaxy. 

Figure \ref{fig:maps_fb_vary} displays the face-on projections of gas surface density and the density-weighted gas temperature at $t$ = 500 Myr for different stellar feedback runs. 
In all runs, hot bubbles are visible between cold, dense spiral arms. However, the thermal and morphological feature of ISM is significantly affected by the stellar feedback strength.
The region where surface density is above 10$^2\msun\,pc^{-2}$ is significantly reduced in the higher stellar feedback runs.
Thermal feedback occurs in the core of the dense and cold regions; the stronger stellar feedback efficiently disrupts these dense ISM region.
The amount of thermal energy determines the balance between formation and destruction of the dense gas clump and directly affects the amount of star formation.

Figure \ref{fig:maps_bulge_vary} presents the edge-on projections of bulge and disk stars in the disk with different $\mathrm{B/T}$ ratio and their face-on projections of gas surface density at $t$ = 500 Myr. 
One can note that the morphological structure of ISM is significantly affected by the $\mathrm{B/T}$ ratio.
The disk-dominated galaxy has fewer but denser arms (see also Figure~\ref{fig:spiral-arm}).
The bulge-dominated galaxy presents strong fragmentation in the ISM, exhibiting a higher number of massive gas clumps than the bulge-dominated galaxy (see also Figure~\ref{fig:clump}).
For the bulge-dominated galaxy, the deep gravitational potential in the center enhances the tidal forces and the ISM experiences a strong shear.
The massive bulge also reduces the turbulence of the ISM and achieves gravitational stability (high Toomre $Q$ parameter), which reduces star formation in the disk and consistent with the idea of morphological quenching \citep{Gensior2020}.

\section{Spatial distribution of gas and stars}
\label{sec:spatial}
To further investigate the differences in the spatial distribution of the gas and stars between the simulations presented in Figure~\ref{fig:maps_fb_vary} and \ref{fig:maps_bulge_vary}, we analyze the properties of spiral arms (Section~\ref{subsec:spiral}) and massive gas clumps (Section~\ref{subsec:clump}) in the galaxies. 
Next, in Section~\ref{subsec:spatial_psd}, we measure the spatial clustering of the gas and stars (grouped into age bins) highlighting how the stellar feedback strength and the morphology of the galaxy influence both the clustering of star formation and the spatial distribution of stars.

\subsection{Spiral arm properties}
\label{subsec:spiral}
We measure the amplitude and number of spiral arms following the procedure described in \cite{Yu2018}. 
Specifically, we perform the Fourier decomposition up to mode $m=$ 6 and neglect higher order modes. 
We implement the Fourier fitting based on the following equation:
\begin{equation}
    I(r,\phi)= I_0(r) + \sum\limits_{m=1}^{6} I_m(r)\,{\rm cos}(m\phi+\phi_m), 
\end{equation}
where $I(r,\phi)$ is the azimuthal profile at a radius of $r$ as a function of angle $\phi$, $I_0(r)$ is the azimuthally averaged intensity, $I_m (r)$ is the amplitude of cosine, and $\phi_m$ is the phase angle. 
We measure $I_m$ at fixed radius $r= 4$ kpc (where the bulges end).
We define the spiral arm amplitude, $I_{\rm tot}=\{\sum\limits_{i=1}^{3} I^2_{m_i}\}^{1/2}$, and count the amplitude-weighted mode of spiral arms, $N_{\rm spiral}=\sum\limits_{i=1}^{3} ( I^2_{m_i}{m_i})/ I^2_{\rm tot}$, 
where $m_{i:i\in\{1,2,3\}}$ are the first three dominant Fourier modes of the spiral arm.

Figure~\ref{fig:spiral-arm} presents the spiral arm strength ($ I_{\rm tot}$) and amplitude-weighted number of spiral arms ($N_{\rm spiral}$) in the galactic disk as a function of feedback strength and $\mathrm{B/T}$ ratio.
The error bars indicate the 16$^{\rm th}$ to 84$^{\rm th}$ percentiles obtained from the 10 simulations with the different random seeds.
The spiral amplitude strictly decreases with the feedback strength, however, the number of the spiral arms stays constant, 4, independent of the feedback strength.
This indicates that SN thermal feedback efficiently suppresses the power of spiral arms without affecting their overall structure.
In contrast, the number of spiral arms depends on the $\mathrm{B/T}$ ratio.
The number of spiral arms increases with $\mathrm{B/T}$ ratio, particularly at low $\mathrm{B/T}$ ratios.
A massive bulge causes a steeper gravitational potential
and increases the angular velocity toward the galactic center. 
Therefore, it increases the azimuthal direction velocity and the shear force tears off the gas clump and spiral arm, enhancing the gravitational stability. 
This quenches star formation within the bulge by preventing the gas from collapsing.

\subsection{Massive clump properties}
\label{subsec:clump}

In order to identify gas clumps, we utilized the {\sc Hop} halo finder algorithm \citep{Eisenstein1998} using gas particles.
We only consider clumps with a radius of less than 0.5 kpc and which are beyond 5 kpc from the galactic center in order to exclude large structures, such as the bulge or spiral arms. 

Figure~\ref{fig:clump} presents the time averaged mass function (left panels), the median number of gas clumps identified (middle panels) and the mass ratio of young-star ($<10$ Myr) in the clumps and the whole galaxy (right panels) for $t=$ 200 -- 1000 Myr.
In the runs that possess many clumps---weak feedback runs or disk-dominated galaxies---the mass function decreases with mass, and the overall slope of the function is approximately $-0.5$. 
A clear dependence of $N_{\rm clump}$ on the strength of feedback and bulge emerges.
Both stronger feedback and a more prominent bulge inhibit the fragmentation of the disk and the formation of gas clumps.
The feedback strength does not significantly affect the slope of the clump mass function, while bulge-dominated galaxies lack massive and low-mass clumps.
This implies that the strong bulge suppresses the formation of a massive clump and also inhibits the survival of a low-mass clump.

As presented in the right panels, the fraction of young stars in clumps, an indicator for the importance of star formation in clumps, is for all runs of the order of $0.1 - 1\%$. 
There is a weak trend of a decreasing ratio with $\mathrm{B/T}$, indicating that the sites of star formation in the galaxy depends on the bulge strength while being invariant to the change of feedback strength.

\subsection{Spatial PSD of the SFH}
\label{subsec:spatial_psd}

\begin{figure*}
\centering
\includegraphics[width=16cm]{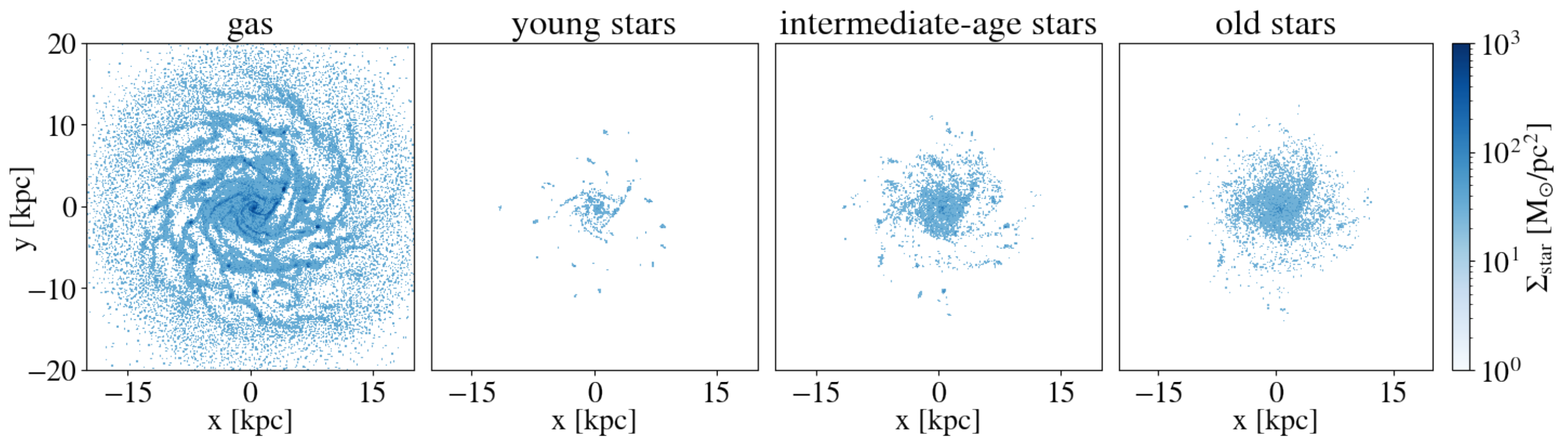}
\caption{
Face-on maps of gas ({\it left most} panel), young stars ($<10$ Myr; {\it 2nd} panel), intermediate-age stars ($10-200$ Myr; {\it 3rd}  panel), and old stars ($>200$ Myr; {\it right most} panel) for {\tt Fiducial} run at $t=$ 500 Myr.
The map of young stars follows dense regions in the gas distribution, displaying a number of small scale structures, such as clumps and spiral arms.
As the age of the stars increases, sizes of star clumps become more dispersed.
See Section~\ref{subsec:spatial_psd} for details.
}
\label{fig:SFH_spatial_example}
\end{figure*}

\begin{figure*}
\centering
\includegraphics[width=18cm]{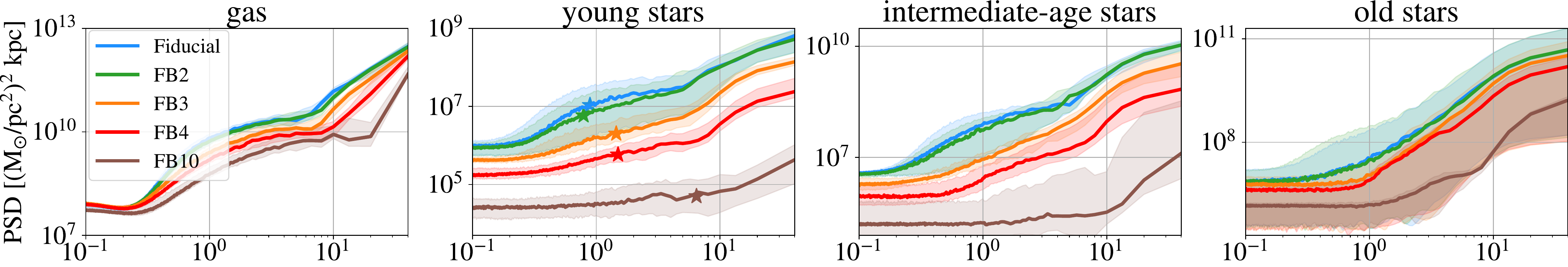}
\includegraphics[width=18cm]{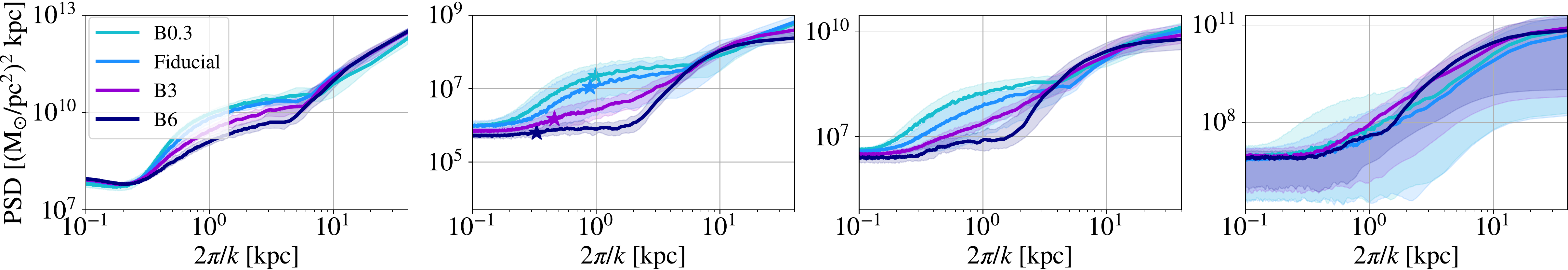}
\caption{The time-averaged spatial PSD of gas and stars (classified by their age) with various feedback strengths ({\it top} panels) and bulge masses ({\it bottom} panels). {\it Left most} panel presents the spatial PSD of gas and the age increases from left to right: young stars ($<10$ Myr; {\it 2nd} panel), intermediate-age stars ($10-200$ Myr; {\it 3rd}  panel), and old stars ($>200$ Myr; {\it right most} panel).
The solid lines indicate median PSD of $t$ = 10 -- 1000 Myr. The shaded regions show 16th to 84th percentiles in all panels. 
The star marks presented in the {\it 2nd} panels indicate the location of break of PSD ($\tau_{\rm break}$).
The strong feedback suppresses the surface density fluctuation on all spatial scales, while the dependency on bulge mass emerges on scales of about 1 kpc. See Section~\ref{subsec:spatial_psd} for details. }
\label{fig:sPSD}
\end{figure*}

In Figure \ref{fig:SFH_spatial_example}, we show the $t$= 500Myr face-on maps of the gas and stars in different ages for {\tt Fiducial} run.
We classify the stars by their ages with following criteria\footnote{We only consider newly formed stars and exclude the stars in the initial condition}:
\begin{itemize}
    \item young stars: $<10$ Myr (roughly probing H$\alpha$-based SFRs)
    \item intermediate-aged stars: $10-200$ Myr (roughly probing UV-based SFRs)
    \item old stars: $>$ 200 Myr
\end{itemize}

Considering large scales, all maps of gas and stars of different ages exhibit  similar centrally concentrated profiles.
The distribution of young stars are identical to the dense region in the maps of gas, showing many small scale structures, such as clumps and spiral arm.
While younger stars show significant clustering, older stars are distributed more uniformly, indicating that young star clusters are quickly dispersed:
\begin{equation}
t_{\rm relax} \sim1{\rm Myr}\left({\frac{R}{10 \rm {pc}}} \right)^{3/2}\left(\frac{N}{10^3}\right)^{1/2}\left(\frac{{\rm ln}N}{{\rm ln} 10^3}\right)^{-1/2}
\end{equation}
where $R$ is the radius of star clusters and $N$ is the number of particles in the system.
Since $t_{\rm relax}$ is on the $\sim$1 Myr, the stars in the clump older than 100 Myr experiences hundreds of $t_{\rm relax}$, relaxed and evaporate from the birth regions.

Figure~\ref{fig:sPSD} displays the time-averaged spatial PSDs of gas and stars of different ages and shows the impact of feedback strength and morphology.
The solid line represents the median spatial PSD for $t=$ 100 -- 1000 Myr with 10 Myr timesteps (i.e., 90 snapshots), and the shaded regions display the 16th and 84th percentiles.
Three features emerge in all plots: (1) the constant noise on small scale ($<$ 300 pc); (2) small bump around $\sim$ 1 kpc scale, which associates the clumps and spiral arm; and (3) the steady increase at $\gtrsim$ 10 kpc which corresponds to the size of the disk.
The spatial PSD of gas exhibits a strong correlation at $>$ 10 kpc scale, while that of stars become flattened at 10 kpc, which results from the confined range of stellar distribution; gas $\gtrsim$ 30 kpc and star $\sim$ 10 kpc. 
Due to the absence of small scale structure in the distribution of old stars, the 1 kpc bump does not exist for the old stars.

The stronger stellar feedback suppresses the power on all scales and also inhibits the formation of 1 kpc bump, which corresponds to spiral arms and clumps (see Figure~\ref{fig:spiral-arm} and ~\ref{fig:clump}).
In the distribution of the stars, the 1 kpc bump is hardly observed in the higher feedback strength runs than the {\tt FB2} run.
We can observe that for the plateau at the 1 kpc scale for the gas, the location of the break---where the slope changes to constant, shown as {\it star} mark---slightly moves to the longer scale due to the larger feedback strength, which efficiently destroys dense clumps and causes diffuse spiral arms.

For the bulge strength test, a great convergence can be observed on small and large scales, but large variations in the 0.3 -- 5 kpc range.
As the age of the star increases, the fluctuation power on the 10 kpc scale increases while the 1 kpc scale fluctuation decreases.
Note that the location of the break moves to the smaller scale.
The huge gravitational potential at the center in the bulge-dominated disk allows the large angular momentum with a large Toomre Q factor and highly inhibits the formation of substructures and tears the dense structure into small scales \citep{Gensior2020}.

\section{How gas cycles through the galaxy and forms stars}
\label{sec:temporal}
In this section, we track individual gas parcels moving inward, outward, and within the simulated galaxies. 
We start by studying the residency time of the gas in the dense and the diffuse phase in Section~\ref{subsec:dense_vs_diffuse}. 
In Section~\ref{subsec:tracking}, we generalize this approach and investigate the evolutionary tracks of the gas parcels. 
Finally, in Section~\ref{subsec:temporal_PSD}, we show how the variability of the SFR---an observable in principle---can be used to infer the residency time of the gas in the dense phase via the measurement of the temporal PSD of the SFH.

\subsection{Residence time in dense versus diffuse regions}
\label{subsec:dense_vs_diffuse}

We now proceed with a detailed analysis on how gas evolves and forms stars, and how the stellar feedback strength and galaxy morphology affect this baryon cycle.
We randomly sample 10$^4$ gas particles (10 percent of the entire gas content) in the galaxy and track them with a time resolution of 1 Myr for the period of $t=750-1000$ Myr. 
We divide the gas states into the diffuse and dense phases, setting the boundary to the star formation threshold density, $\rho_{\rm SF, thres}$, and measure the periods spent in these phases for each gas parcel during the 250 Myr.

\begin{figure*}
\centering
\includegraphics[width=18cm]{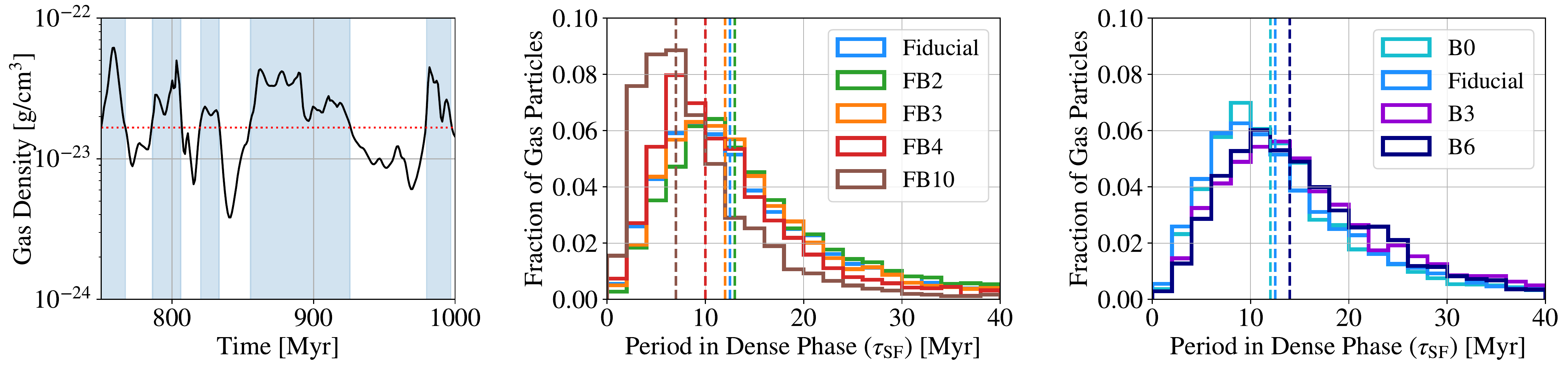}
\includegraphics[width=18cm]{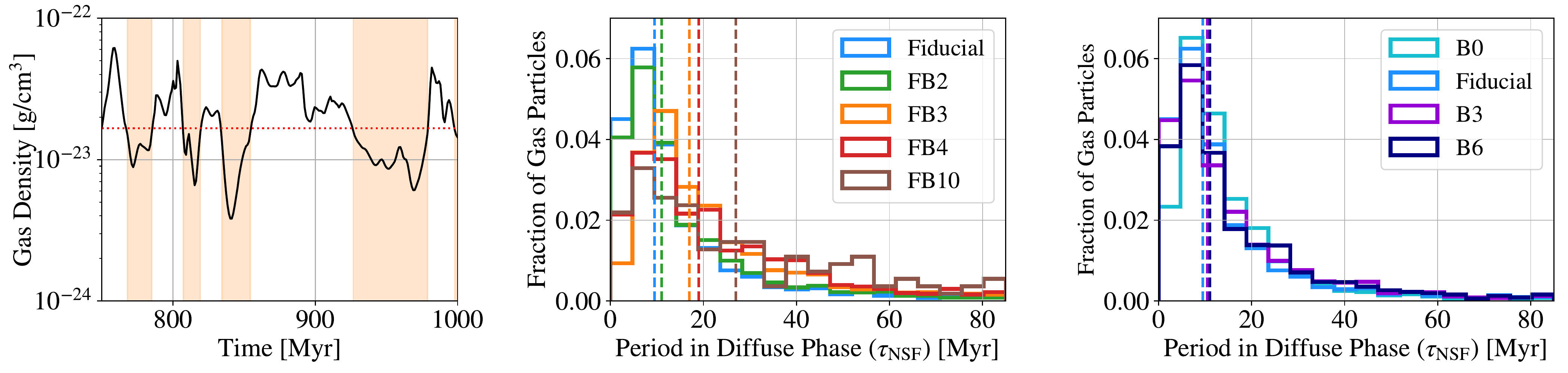}
\caption{
Measurement of the period of gas in the dense ({\it top} panels) and diffuse ({\it bottom} panels) ISM phase. An example of gas density history during $t=750-1000$ Myr is plotted in the {\it left} panels. The horizontal {\it red} lines mark the threshold density ($\rho_{\rm SF, thres}$), the boundary between the dense and diffuse phases. The shaded {\it blue} and {\it orange} regions highlight the period when the particle resides in the dense and diffuse phase, respectively. The {\it middle} and {\it right} panels present the histograms of the period in the dense ({\it top}) and diffuse ({\it bottom}) phase as a function of feedback and bulge strength, respectively. The vertical lines represent the median of the period in the dense (or diffuse) phase. Stronger feedback reduces the lifetime of gas in the dense star-forming state and causes the gas to reside longer in the diffuse phase. The period of the gas in the diffuse or dense phase does not significantly depend on the bulge mass. See Section~\ref{subsec:dense_vs_diffuse} for details.}
\label{fig:gas_density_evolution}
\end{figure*}

The left panels of Figure~\ref{fig:gas_density_evolution} present an example of a gas density history for $t=750-1000$ Myr. The middle and right panels show the histograms of the times spent in the dense ({\it top} panels) and diffuse (bottom panels) phase, as a function of stellar feedback strength and morphology, respectively.
The histograms present clear trends in both dense and diffuse phases with different strengths of feedback.
The period in the dense phase decreases in the runs with the stronger feedback energy, and the reverse is true for the diffuse phase.
Injection of higher feedback energy increases the turbulence in the ISM and inhibits the gas from residing in the dense phase.
Contrarily, the effect of the bulge strength is only mild: stronger bulge slightly increases the duration of dense phases and does not have any systematic effect on the duration of diffuse phase.

\subsection{Tracking individual gas parcels}
\label{subsec:tracking}

\begin{figure*}
\centering
\includegraphics[width=15cm]{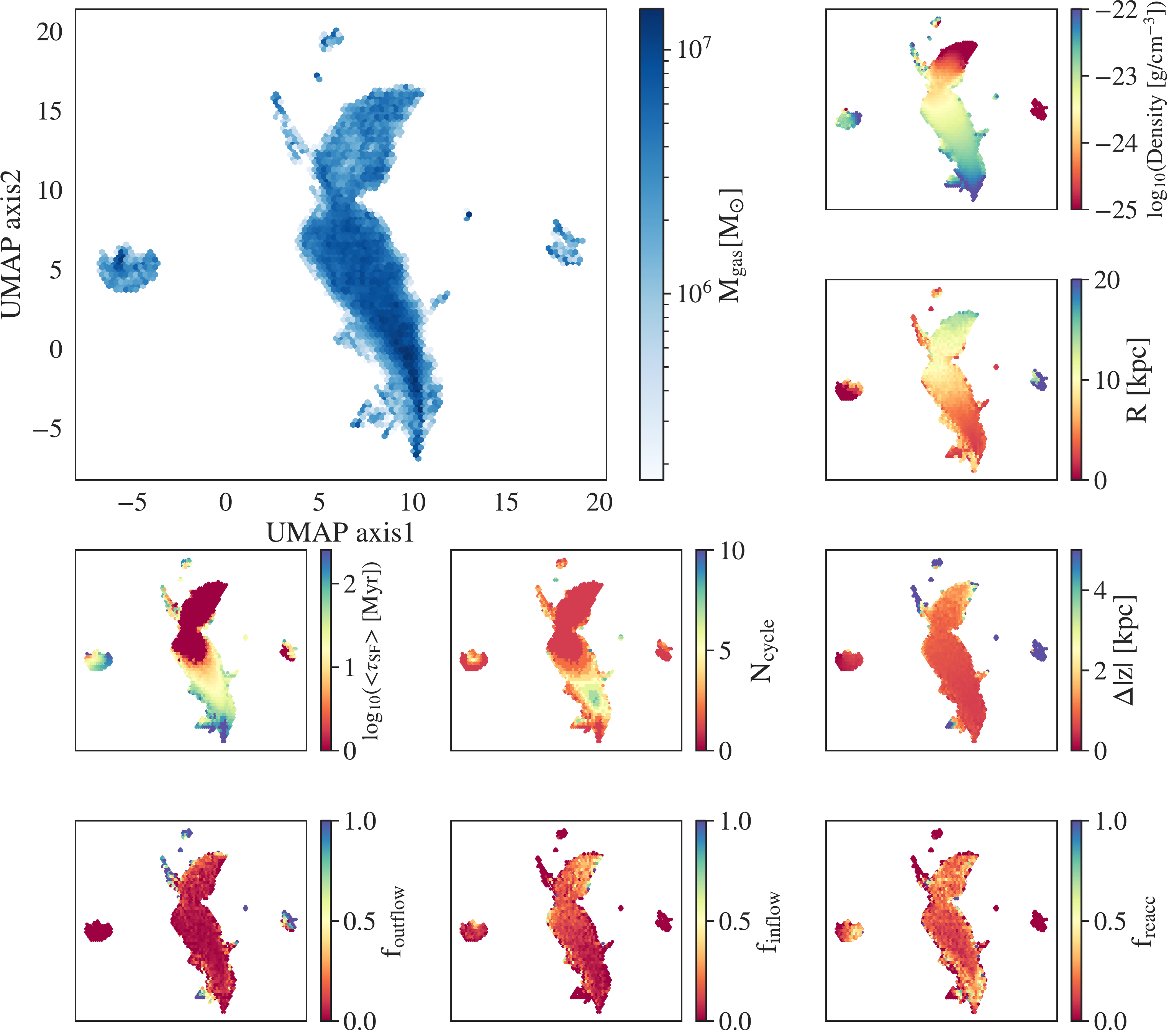}
\caption{
Gas cycle classification using the {\sc Umap} algorithm.
We track 8$\times10^4$ gas particles for $t=$ 750 -- 1000 Myr and use their density histories as the input to the {\sc Umap} algorithm.
{\it Top left} panel presents the result of the {\sc Umap} clustering with 8$\times10^4$ data points on the 2-dimensional {\sc Umap} axes. 
Each point in the maps represents a trajectory of a gas particle density over 250 Myr.
We then plot the projections on the {\sc Umap} space colored with the median density ({\it top right} panel), the median distance from the galactic center ($R$; {\it 2nd row right} panel), average time period in star-forming phase, ($\tau_{\rm SF}$; {\it 3rd row left most} panel), number of cycle across star-forming and non-star-forming phase ($N_{\rm cycle}$; {\it 3rd row middle} panel), height difference ($\Delta|z|=|z|_{\rm max}-|z|_{\rm min}$; {\it 3rd row right} panel) for $t=$ 750 -- 1000 Myr, fraction of outflow ($f_{\rm outflow}$; {\it bottom left most} panel), inflow ($f_{\rm inflow}$; {\it bottom middle} bottom middle panel), and re-accreted gas ($f_{\rm reacc}$; {\it bottom right most} panel) in the bin. 
We can interpret what each cluster represents using the physical properties found in this analysis (see Figure~\ref{fig:UMAP-class}). 
See Section~\ref{subsubsec:umap-physics} for details.}
\label{fig:umap-physics}
\end{figure*}

Now we focus on the gas cycle within the galaxy, tracking the density history of gas particles.
The tracks of gas densities show a significant diversity as gas transits between dense and diffuse phases multiple times with a range of pathways, which is challenging to classify.
Therefore, we employ the {\sc Umap} algorithm \citep{McInnes2018} for classifying the evolutionary histories of individual gas particles.
{\sc Umap}, which stands for `Uniform Manifold Approximation and Projection', categorizes given samples by learning their geometrical structures on an $N$-dimensional manifold using the simplicial complexes concept.
{\sc Umap} illustrates the resulting clusters on the reduced dimensional axes of the manifold, suggested by the algorithm\footnote{For more information about the {\sc Umap} algorithm, see \url{https://umap-learn.readthedocs.io/en/latest/index.html}.}.

We extract the gas density histories for 250 Myr---roughly the dynamical time for a Milky Way-mass galaxy---and inserted an additional tag for the gas particles which become stars.
We run the {\sc Umap} algorithm with all of the samples from each setup, the density histories of 10$^4$ gas particles for eight setups (8$\times10^4$ gas particles) in order to apply a single kind of classification criteria.
When we consider a specific density history of gas particles, which is regarded as a continuous function in time, there exist many other gas particles which have similar evolutionary trends. 
These density history functions are locally connected in the manifold and become a group in the {\sc Umap} analysis.

\subsubsection{Examining of physical quantities of various gas cycles in {\sc Umap} classification}
\label{subsubsec:umap-physics}
Figure~\ref{fig:umap-physics} presents the result of the {\sc Umap} clustering. 
Each data point represents a single gas particle carrying the entire evolutionary information in density for 250 Myr and the hexagonal histograms of $8\times10^4$ data points are displayed with the 2-dimensional axes suggested by the {\sc Umap} algorithm.
Many clusters in the {\sc Umap} projections are presented, including a large cluster located in the center and several small clusters.
Gas particles in the disk have many other gas particles which are similar or continuously varying history in density in the disk; therefore, the disk gas becomes the huge cluster in the {\sc Umap} classification.
On the other hand, the gas that turns into stars or outflowing gas will exhibit clear trends---i.e., significant change in density. These gas can then be classified as distinct small groups.

To understand the characteristics of this {\sc Umap} projection, we investigate the distribution of the following physical quantities: the median density during the 250 Myr, the median distance from the galactic center ($R$), the time in the star-forming state ($\tau_{\rm SF}$), the number of cycles across the density threshold ($N_{\rm cycle}$), deviations of maximum and minimum heights from the galactic plane ($\Delta|z|$), the fraction of outflow ($f_{\rm outflow}$), inflow ($f_{\rm inflow}$), and re-accreted ($f_{\rm reacc}$) material, and plot these quantities in the 2-dimensional {\sc Umap} projection.

To classify the outflow, inflow, and re-accreted gas, we check whether the gas is `in' or `out' of the disk at initial ($t$ = 750 Myr) and final ($t$ = 1 Gyr) timestep, and classify the gas using the following criteria:
(i) outflow gas: `in' $\rightarrow$ `out',
(ii) inflow gas: `out' $\rightarrow$ `in',
(iii) re-accretion gas: `in' $\rightarrow$ `out' $\rightarrow$ `in'.\footnote{Note that in the complete classification, `in' $\rightarrow$ `in' and `out' $\rightarrow$ `out' cases exist, which we do not plot.}
For this classification, we define the disk as the region where the radial distance from the galactic center is less than 15 kpc ($\sim 5\,r_{\rm d}$) and the height from the galactic plane is less than 1.5 kpc ($\sim 5\,z_{\rm d}$).

\begin{figure*}
\centering
\includegraphics[width=18cm]{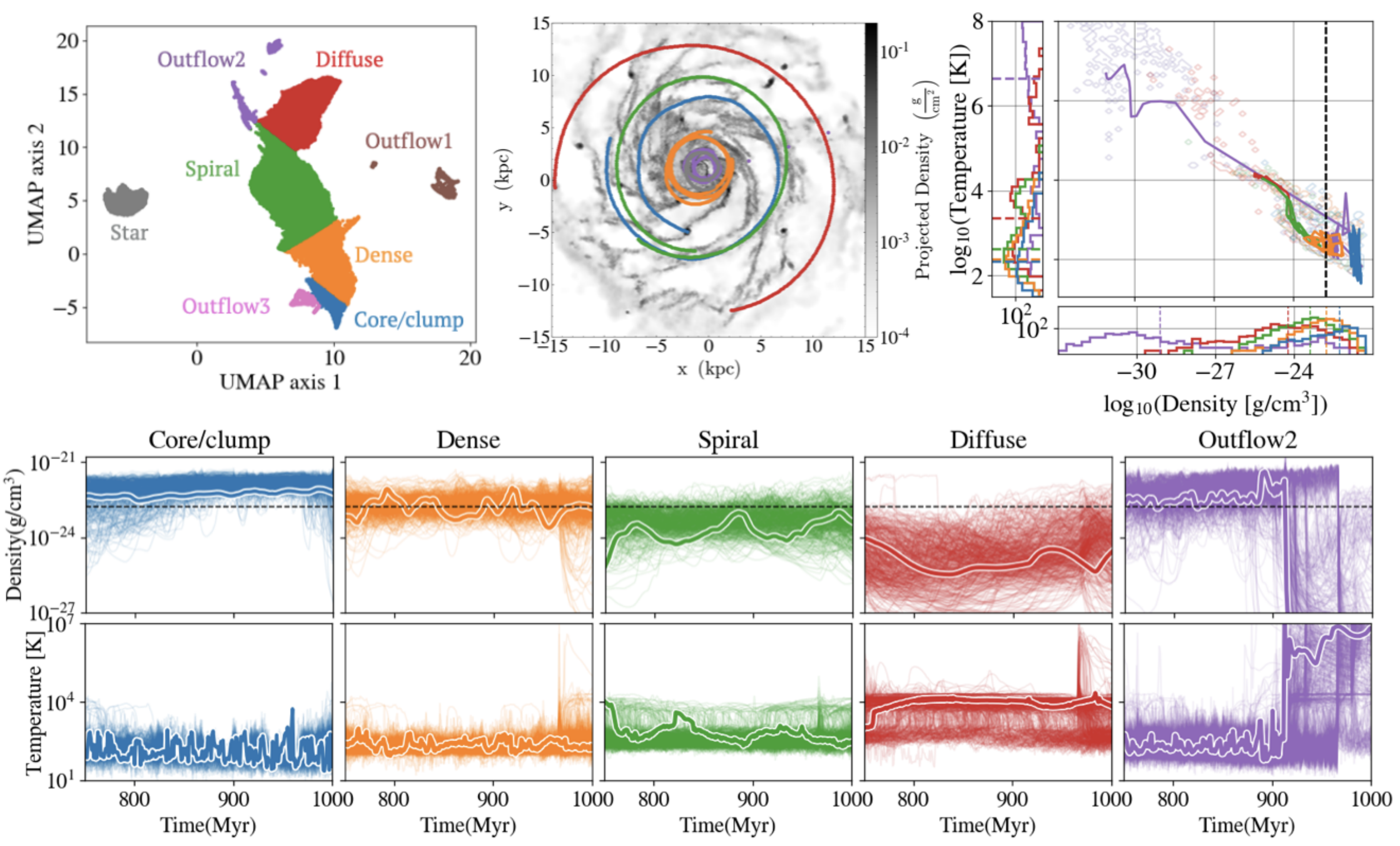}
\caption{
Classification of gas density trajectories based on the correlations in the {\sc Umap} space presented in Figure~\ref{fig:umap-physics}.
We classify gas particles into `Core/clump' ({\it blue}), `Dense' ({\it orange}), `Spiral' ({\it green}), `Diffuse' ({\it red}), `Outflows1, 2, 3' ({\it brown, purple, pink}, respectively), and `Star' ({\it grey}) groups and plot the trajectories of a selected gas particle for each group on the x-y plane ({\it top middle} panel) and the density-temperature plane ({\it top right} panel). 
We display the density ({\it middle} panels) and temperature histories ({\it bottom} panels) for individual particles ({\it thin} lines) of each groups, while the {\it highlighted} lines are the evolutionary histories for the selected gas particle (from {\tt Fiducial} run) presented in {\it top right} panel.
The density threshold line is shown with the {\it black dashed} line. 
See Section~\ref{subsubsec:umap-cycle} for details.}
\label{fig:UMAP-class}
\end{figure*}

\begin{figure*}
\includegraphics[width=15cm]{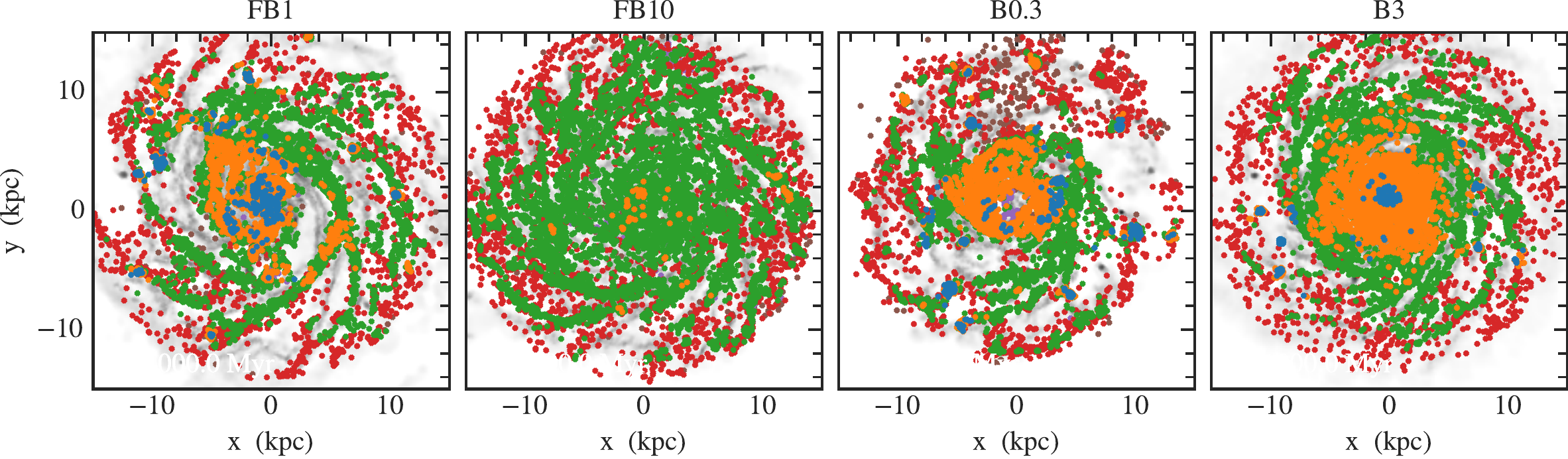}
\includegraphics[width=15cm]{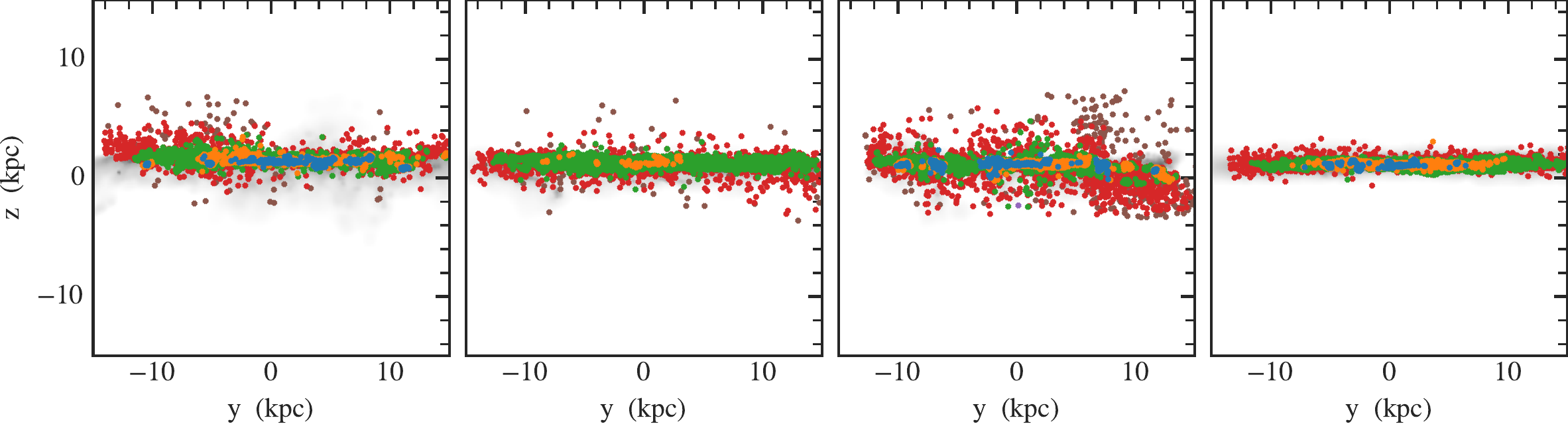}
\includegraphics[width=2.6cm]{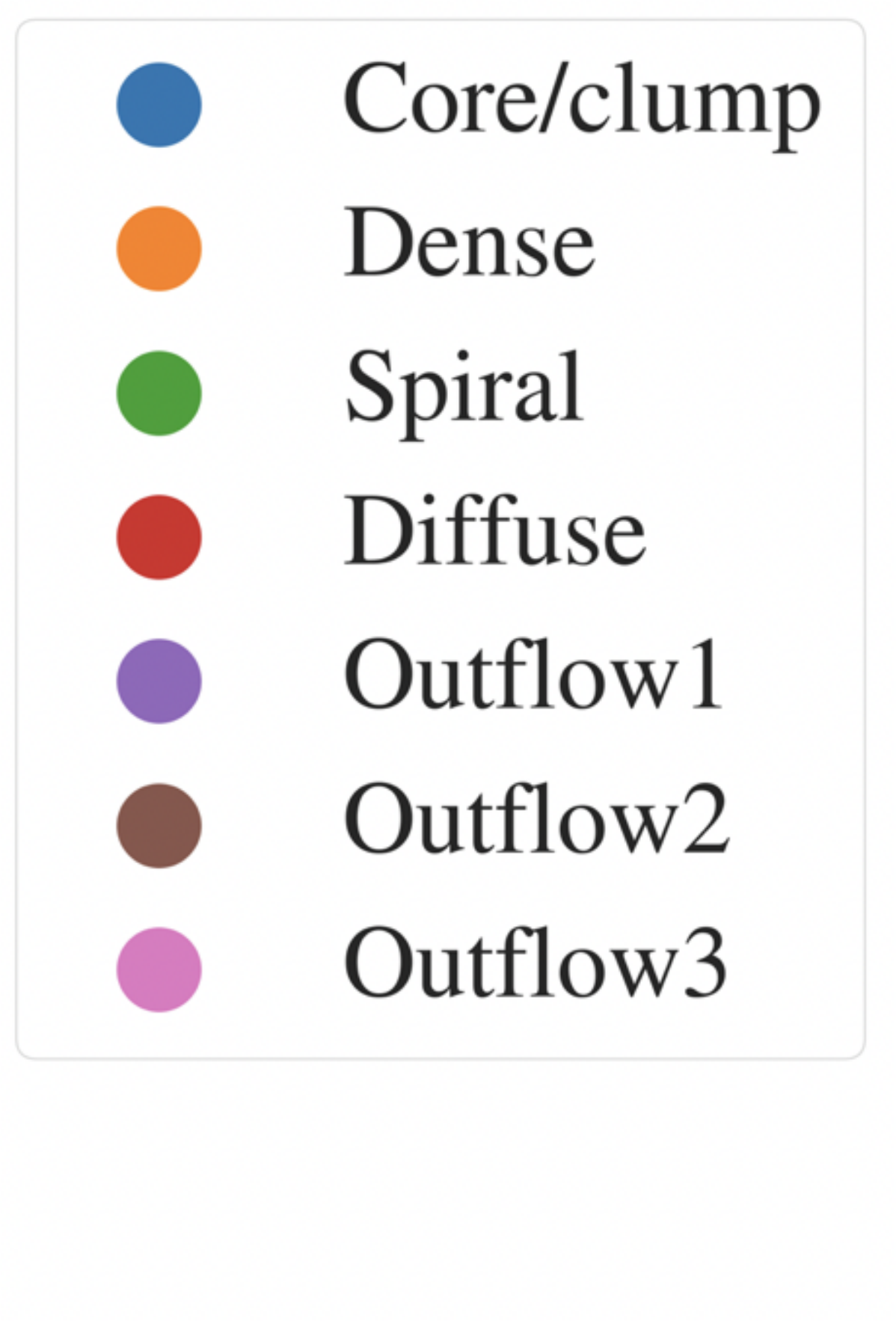}
\caption{
Face-on and edge-on maps of gas particles colored by group classification presented in Figure~\ref{fig:UMAP-class}: `Core/clump' ({\it blue}), `Dense' ({\it orange}), `Spiral' ({\it green}), `Diffuse' ({\it red}), `Outflows1, 2, 3' ({\it brown, purple, pink}, respectively), and particles that turn into stellar particles as `Star' ({\it grey}). 
We plot the face-on ({\it top} panels) and edge-on ({\it bottom} panels) projections of gas particles in different classes at $t=$ 1 Gyr for {\tt Fiducial} ({\it left most} panels), {\tt FB10} ({\it 2nd} panels), {\tt B0.3} ({\it 3rd} panels) and {\tt B3} ({\it right most} panel) run.
Stronger feedback reduces the amount of dense star-forming gas, while increasing the bulge strength leads to a more stable disk with a smoother distribution of gas.
See Section~\ref{subsubsec:umap-fb-bt} for details.
}
\label{fig:umap-proj}
\end{figure*}

\begin{figure}
\centering
\includegraphics[width=8.5cm]{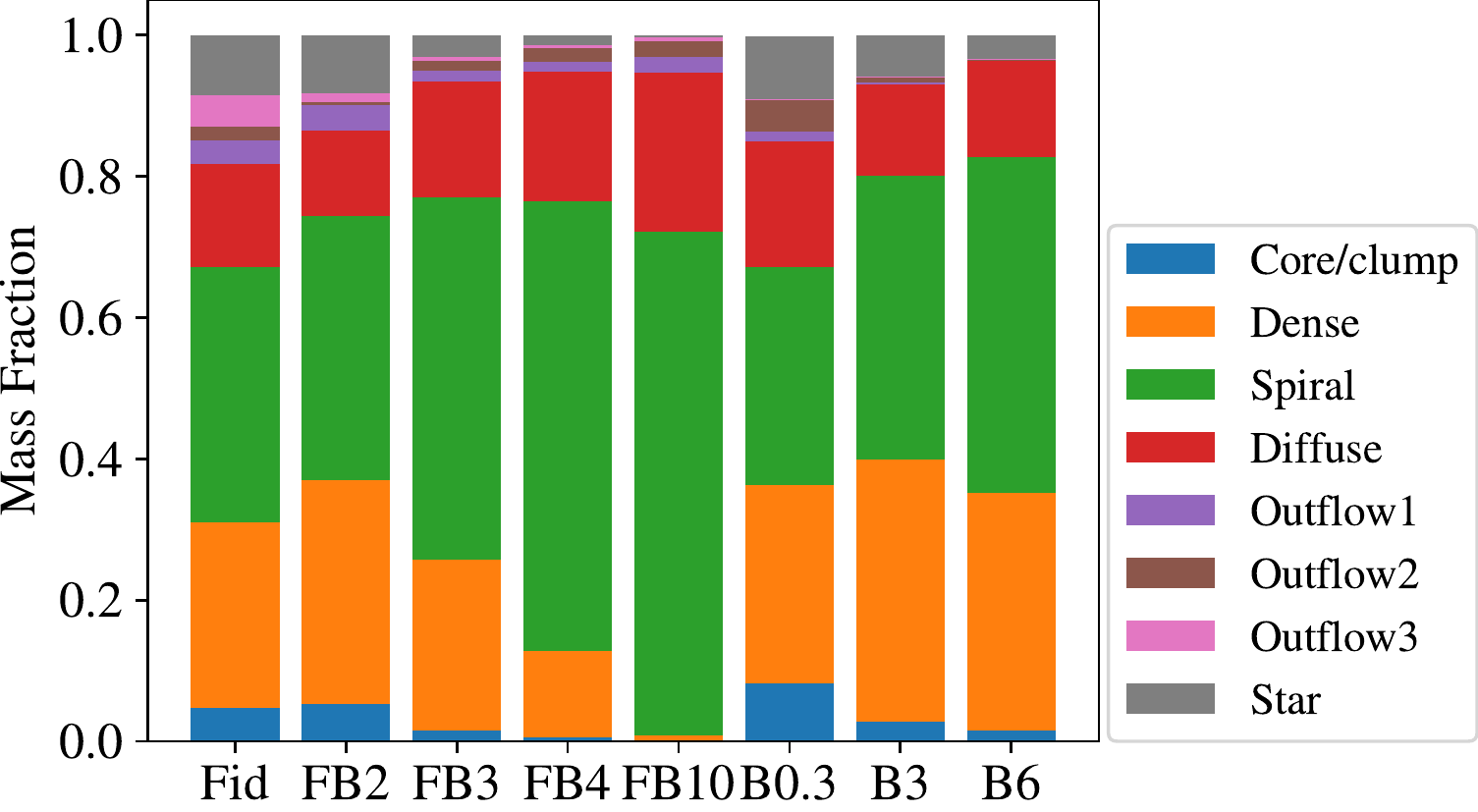}
\caption{
Mass fraction of gas particle trajectories classified in each simulation setup: `Core/clump' ({\it blue}), `Dense' ({\it orange}), `Spiral' ({\it green}), `Diffuse' ({\it red}), `Outflows1, 2, 3' ({\it brown, purple, pink}, respectively), and `Star' ({\it grey}). 
Both stellar feedback and bulge strength significantly affect the distribution of particles between different types of cycles.
See Section~\ref{subsubsec:umap-fb-bt} for details.}
\label{fig:UMAP-fraction}
\end{figure}

First, we find a strong correlation between the density and the distance from the center ($R$): gas in the central region is denser than on the disk outskirts.
This result suggests that gas density tracks imprint information about gas location in the disk.
The number of cycles of gas between star-forming and non-star-forming states, $N_{\rm cycle}$ (also see Section~\ref{subsec:dense_vs_diffuse}), is large ($>5$) in the middle of the main clump, while it is low for the other parts of the diagram.
$\Delta|z|$ and $f_{\rm outflow}$ also show significant correlation across different {\sc Umap} components as regions with large differences in $|z|$ values are associated with gas outflows from the disk.
Comparing the bottom panels reveals that the majority of the gas has been re-accreted. 
This is dependent on the definition of the disk---a cylinder with 5 times both the scale radius and height, but the result indicates that the majority of the gas deviate by five times the scale lengths and re-accrete to the disk.

\subsubsection{Evolution of density, temperature, and position of different gas cycle within galaxies}
\label{subsubsec:umap-cycle}
Based on Figure~\ref{fig:umap-physics}, we categorize 8$\times10^4$ gas particles into six groups: `Core/clump', `Dense', `Spiral', `Diffuse', `Outflow1, 2, 3', and `Star'.
In Figure~\ref{fig:UMAP-class}, the top left panel exhibits our identification of the gas particle groups.
The group colored in grey is the gas particles turning into star particles during the 250 Myr. 
We identify outflow gas based on $f_{\rm outflow}$ projection, and name them as `Outflow1, 2, 3', in order of the time when the outflow occur.
We categorize the remaining large group, which corresponds to the gaseous disk, into `Core/clump', `Dense', `Spiral', `Diffuse' groups based on the median density shown in Figure~\ref{fig:umap-physics}.

The two bottom panels in the Figure~\ref{fig:UMAP-class} presents evolution of the density and temperature of individual gas from each group for $t$ = 750 -- 1000 Myr.
The highlighted lines are examples of the evolutionary history from each groups.
Gas density shows significant variations with its range depending on the environment: `Core/Clump' group with 10$^{-23} - 10^{-21} {\rm g\,cm^{-3}}$, `Dense' group with 10$^{-24} - 10^{-22} {\rm g\,cm^{-3}}$,  `Spiral' group with 10$^{-25} - 10^{-23} {\rm g\,cm^{-3}}$, `Diffuse' group with 10$^{-27} - 10^{-24} {\rm g\,cm^{-3}}$.
The gas in denser phases is colder, and vice versa. Therefore, the histories of density and temperature evolves in the opposite direction: the majority of gas in `Core/clump' and `Dense' group resides in the cold phase ($\sim 10^2$ K) while the gas in `Spiral' and `Diffuse' spends more time in the warm phase ($\sim 10^4$ K). The timescale on which gas densities and temperatures change is the shortest in `Core/clump', and becomes longer in the `Dense', `Spiral',  and `Diffuse' groups in that order. This trend reflects longer dynamical and cooling timescales in lower density environments.
In contrast to these steady oscillations, the gas in the outflow group shows a dramatic change in density and temperature with time; it is initially dense and cold but then its density dramatically decreases and temperature increases following an outflow episode.
We can observe that right before the outflow happens, the density of the `Outflow' group reaches that of `Core/clump' gas.
This is consistent with our understanding that the outflow happens in the dense ISM nearby young stars.

For `Core/clump', the oscillation period is around 10 Myr, which is the crossing time of star cluster,\footnote{In our simulations, the internal dynamics of such clumps is not resolved and therefore their lifetimes are expected to be even shorter.}
\begin{equation}
\begin{aligned}
\tau_{\rm clump}\sim 10\,{\rm Myr}\left(\frac{M}{10^{8}\msun}\right)^{-1/2}\left(\frac{R}{0.4\, {\rm kpc}}\right)^{3/2}
\end{aligned}
\end{equation}

The gas in the outskirts enters and exits a spiral arm causing density oscillations, whereas the cycles in `Core/Clump' group are unaffected by the spiral arm. 
The period of density oscillation is primarily determined by the length of time of the gas between successive passages of spiral arms:
\begin{equation}
\begin{aligned}
\tau_{\rm arm} 
\sim 110\,{\rm Myr}\left(\frac{R}{7 \,{\rm kpc}}\right)\left(\frac{N_{\rm arm}}{4}\right)^{-1}\left({\frac{v_{\rm rel}}{100\, {\rm km\,s^{-1}}}}\right)^{-1}
\end{aligned}
\end{equation}
where $R$ is orbital radius, $N_{\rm arm}$ is number of spiral arm (we choose 4, see Figure~\ref{fig:spiral-arm}) in the disk and $v_{\rm rel}$ is the relative velocity of the gas and spiral arm pattern $v_{\rm rel}=|v_{\rm gas}-v_{\rm pattern}|$ \citep[see also][]{Semenov2017}.\footnote{It is difficult to quantify the velocity of the spiral arms due to their short lifetime ($\sim$ 100 Myr), but the average period can be well explained by the relative movement of gas against the spiral arm. 
The quantification of the density wave of spiral arm is possible using spatio-temporal PSD-SFR (Shin et al. in prep).}
Based on the orbital radius shown in the {\it top middle} panel, the equation explains the density oscillation period in `Dense' ($\sim$ 50 Myr), `spiral' ($\sim$ 110 Myr) and `Diffuse' ($\gtrsim$ 240 Myr).\footnote{Note that we track the density oscillation for 250 Myr; the oscillation periods larger than this scale are not fully measured in this work.}

We present the one and two-dimensional histogram and the median density and temperature for the $t=1$ Gyr snapshot and the trajectories of gas particles in different groups (the same particles presented in the bottom panels) on the density-temperature space.
The `Core/clump' and `Outflow1' gas oscillates in the cold and dense phase and the `Outflow1' gas moves to the hot and diffuse phase.
The `Dense' gas oscillates across the star formation density threshold; note that this group shows the highest $N_{\rm cycle}$ in Figure \ref{fig:umap-physics}.
The gas from `Spiral' and `Diffuse' groups spends the majority of time in the diffuse phase below the star formation threshold.

\begin{figure*}
\centering
\includegraphics[width=15cm]{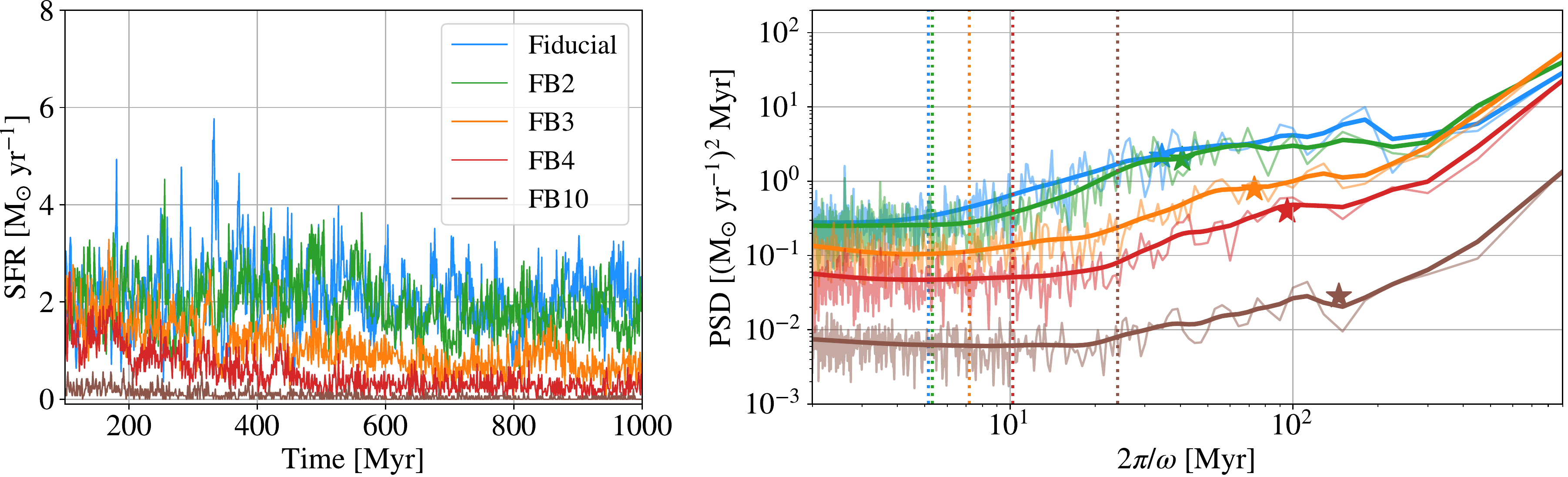}
\includegraphics[width=15cm]{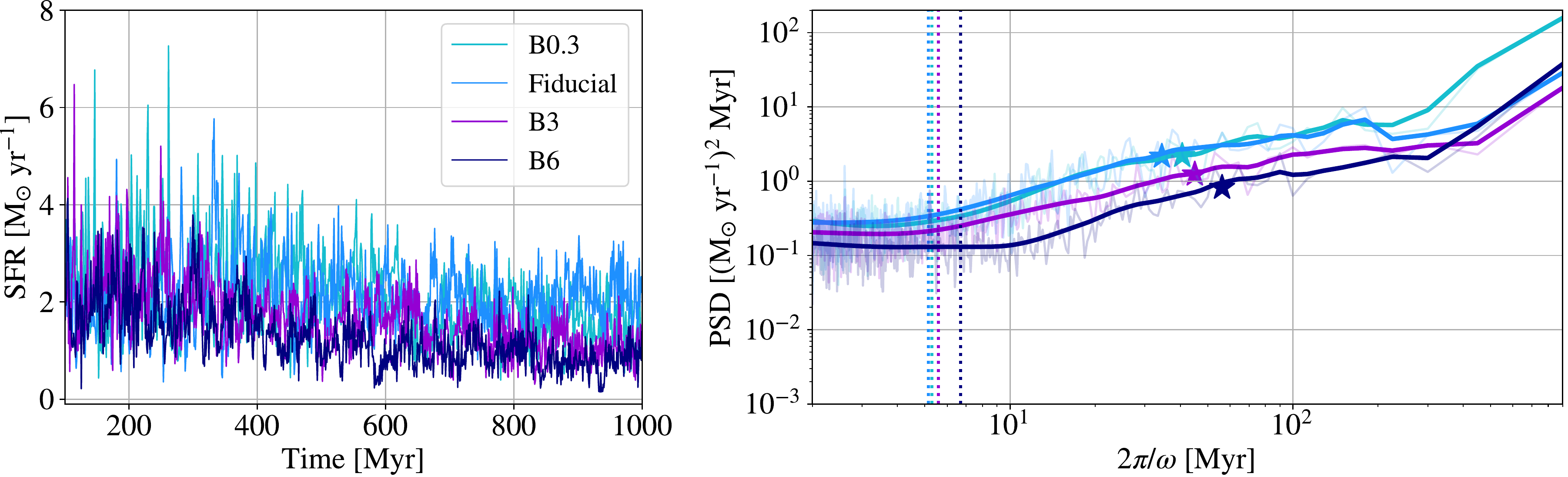}
\caption{Star formation histories (SFHs; {\it left} panels) and the corresponding temporal power spectral densities (PSDs; {\it right} panels) of the simulations with varying stellar feedback strengths ({\it top} panels) and with varying bulge strengths ({\it bottom} panels).
We plot one individual SFH in the {\it left} panels, while we present the median PSDs ({\it thin} lines) and their fitting lines ({\it thick} lines) for all 10 runs with different random seeds of a given setup in the {\it right} panels. 
The break of PSD ($\tau_{\rm break}$) is indicated with star marks on the {\it right} panels, which is calculated by Extended Regulator model \citep[][see Appendix~\ref{appsec:psd_fitting}]{Tacchella2020}.
Stronger stellar feedback leads to a decrease of the overall normalization of the PSD (caused by the lower average SFRs) and an increase in the star formation correlation timescale ($\tau_{\rm break}$).
However, the bulge strength does not change $\tau_{\rm break}$ significantly.
We present the calculated the probing limits of the fluctuation timescale with {\it vertical dotted} lines (see Appendix~\ref{appsec:resolution}). 
The flattening of the PSDs toward short fluctuation timescales is affected by resolution effects and should be interpreted with caution.
See Section~\ref{subsec:temporal_PSD} for details.}
\label{fig:sfh_psd}
\end{figure*}

\subsubsection{The dependence of gas cycle on the strength of stellar feedback and $B/T$ of galaxies}
\label{subsubsec:umap-fb-bt}
Figure~\ref{fig:umap-proj} shows the projections of gas colored by their class for each setup.
The spatial distribution of gas with different evolution histories shows clear trends.
The gas from `Core/clump' group resides in the galactic center or clump regions and is spatially concentrated. The distribution of gas becomes progressively more volume-filling for `Dense,' `Spiral,' and `Diffuse' groups.
It is remarkable that {\sc Umap} classification does not rely on any spacial information about the gas, however, the groups classified by gas density history can differentiate gas by its environment.

{\tt Fiducial} run has an abundant amount of `Core/clump' gas in the galactic disk while {\tt FB10} has much less cold and dense gas.
In the disk-dominated galaxy ({\tt B0.3}), {\it blue} gas particles reside in the outskirt region rather than in the central region, while they are in the core region for the bulge-dominated galaxy ({\tt B3}).
This implies that the bulge mass significantly affects the spatial distribution of dense gas clumps.
The scattered distribution in the vertical direction of diffuse gas in {\tt B0.3} is due to the active star-forming activities, while the gas in {\tt B3} is highly stable because of the morphologically quenched star formation.

Figure~\ref{fig:UMAP-fraction} presents the mass fraction of star particles identified as each group in each simulation setup.  
More than 80 percent of gas resides in the disk (`Core/clump', `Dense', `Spiral' and `Diffuse') during the entire 250 Myr, and less than 20 percent becomes stars or outflow gas.
Also, 40 percent of gas is in the non-star-forming state (`Spiral' and `Diffuse') in for 250 Myr.
Notably, the fractions of `Star' and `Outflows' are proportional to that of `Core/clump.' 
The higher feedback runs contain less `Core/clump' and `Dense' gas, whereas the relative fraction of the `Spiral' and `Diffuse' gas groups increases.
In addition, the bulge strength reduces the fraction of `Core/clump' gas in the galaxy by a significant amount.
Both the stellar feedback and bulge strength have a significant impact on the distribution of gas between different {\sc Umap} groups, implying that both have significant effect on the baryon cycle in galaxies. 

\subsection{Star formation variability as a probe for the baryon cycle} 
\label{subsec:temporal_PSD}

\begin{figure*}
\centering
\includegraphics[width=16cm]{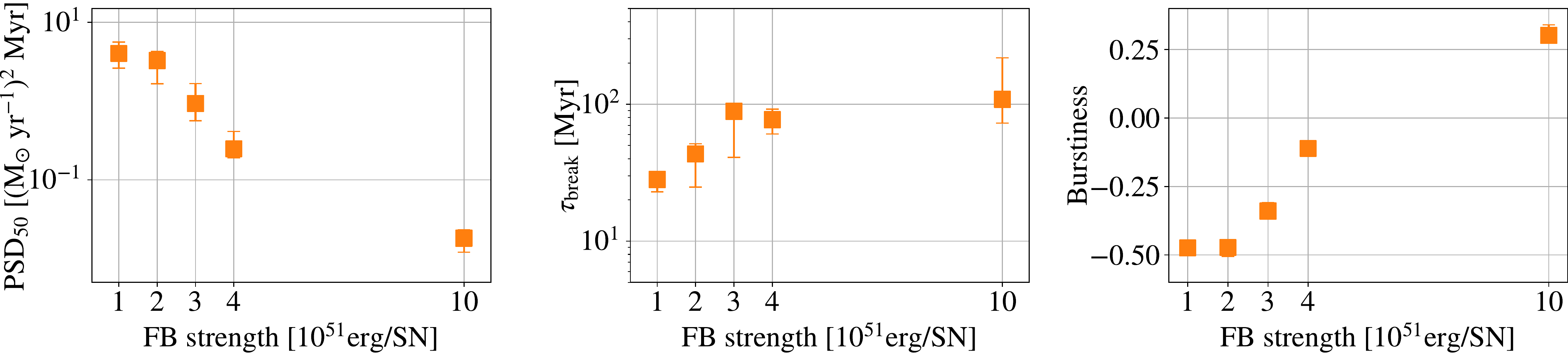}
\includegraphics[width=16cm]{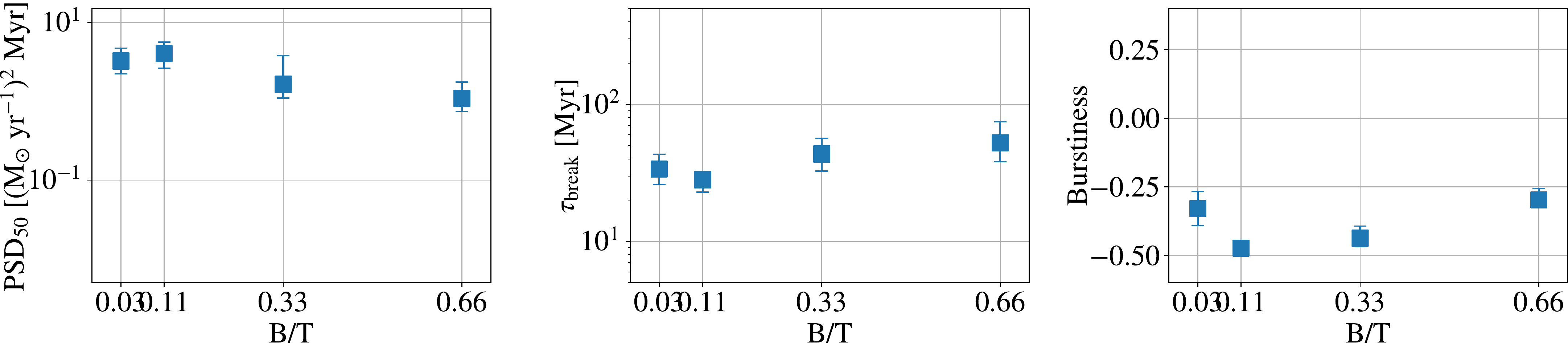}
\caption{
Dependence of the temporal PSD of the SFH on stellar feedback strength and bulge strength. We plot the PSD amplitude at a timescale of 50 Myr (PSD$_{\rm 50}$; {\it left} panels), the timescale of the break of the PSD ($\tau_{\rm  break}$; {\it middle} panels) and burstiness parameter (see Equation~\ref{eq:B}; {\it right} panels)
as a function of stellar feedback strength ({\it top} panels) and $\mathrm{B/T}$ ratio ({\it bottom} panels). 
Increasing feedback strength leads to more bursty star formation and longer correlation timescale ($\tau_{\rm break}$) of the SFH. 
A similar trends are observed for the higher $\mathrm{B/T}$ ratio, however, the dependence is less than the feedback effect.
See Section~\ref{subsec:temporal_PSD} for details.}
\label{fig:psd-amplitude}
\end{figure*}

Figure~\ref{fig:sfh_psd} presents individual SFHs and the median PSDs of all ten runs of each setup.
We compute each PSD of SFH from all ten runs for each setup and calculate the median PSDs of the ten PSDs.
During the first 100 Myr, an initial starburst occurs as the initial conditions used in these simulations are slightly out of equilibrium. The SFH stabilizes after roughly a dynamical timescale. Therefore, we exclude the first 100 Myr from our analysis.

The dependence of PSDs on feedback and bulge strength is further quantified in Figure~\ref{fig:psd-amplitude}: we plot the PSD amplitudes at a timescale of 50 Myr, PSD break ($\tau_{\rm break}$), and burstiness of SFH for the different setups.
$\tau_{\rm break}$ is estimated by fitting the Extended Regulator model studied by \citet[][also see Appendix \ref{appsec:resolution}]{Tacchella2020}. The PSD breaks are marked with a {\it star} in Figure~\ref{fig:sfh_psd}.
We define the burstiness parameter as follows:
\begin{equation}
B=\frac{\sigma/\mu-1}{\sigma/\mu+1}
\label{eq:B}
\end{equation}
where $\sigma$ is the standard deviation of SFH, and $\mu$ is the mean value of SFH \citep{Goh+2008, Caplar+Tacchella2019}. $B=1$ corresponds to a maximally bursty signal ($\sigma/\mu\gg$1) and $B=-1$ is a constant signal ($\mu/\sigma\gg$1).

The temporal PSD function shows similar feature to the spatial PSD shown in Fig~\ref{fig:sPSD}: the power increases on the scale of 10 -- 100 Myr, reaching the plateau at $\sim$ 100 Myr, and is again growing above that fluctuation scale.
This is not too surprising since one expects that larger temporal scales couple to larger spatial scales.
\cite{Tacchella2020} elucidated the SFH variability in galaxy using Extended Regulator model, which describes the feature of SFH PSD in the 1 Myr -- 10 Gyr range (see Appendix~\ref{appsec:psd_fitting}).
The SFH PSD of our simulations are consistent with their model in 10 -- 1000 Myr range: the small scale bumps around the timescale of 10 Myr are induced by the dynamical process and the formation/disruption of GMCs, while the features on the $>$ 200 Myr timescale are explained by the galactic inflow from the circumgalactic fountain \citep{Tacchella2020} or the steady gas consumption which leads to a steady decrease of SFR on global gas depletion timescale \citep{Semenov2017}.
On timescale below 10 Myr, the simulations flatten due to numerical resolutions.
We discuss the time scale of the white noise (presented with vertical lines) in SFH PSD in Appendix~\ref{appsec:resolution}.

The power decreases in the stronger feedback runs at all scales, which is consistent with the idea that stronger feedback leads to lower overall SFRs. 
The amplitude of the fluctuation scale of 50 Myr decreases by a factor of 100 as the feedback strength increases by a factor of 10. 
Note, however, that this decrease of PSD amplitude is mainly due to the suppression of total SFR in runs with stronger feedback; the \emph{relative} variability of SFR is, in fact, increasing with stronger feedback. 
Indeed, the PSD slope decreases with stronger feedback around a timescale of 10 -- 100 Myr, that correspond to the increase of SFR burstiness, which is consistent with the shorter period of star formation regions in the stronger feedback.
Similar to the spatial PSD, the break of temporal PSD ($\tau_{\rm break}$), around a timescale of 10 -- 100 Myr, increases with stronger feedback energy.
Moreover, the noise line, calculated by Equation~\ref{eq:t_res} based on \citet[][see Appendix~\ref{appsec:resolution}]{Iyer2020}, from the short fluctuation timescale extends to the longer scale for the stronger feedback, because of the lack of particles in time bin at lower SFRs. 

The bulge strength also suppresses the SFR, especially on the 10 -- 100 Myr fluctuation timescale, which leads to the suppressed power shown in Figure~\ref{fig:sPSD}.  
The change of the power for 50 Myr fluctuations is by a factor of 3 from $\mathrm{B/T}$ of 0.03 to 0.66.
The burstiness of SFH and the location of PSD break does not significantly change with  $\mathrm{B/T}$, indicating the the SFH is correlated over similar timescales, independent of morphology.

\begin{figure*}
\centering
\includegraphics[width=0.7\textwidth]{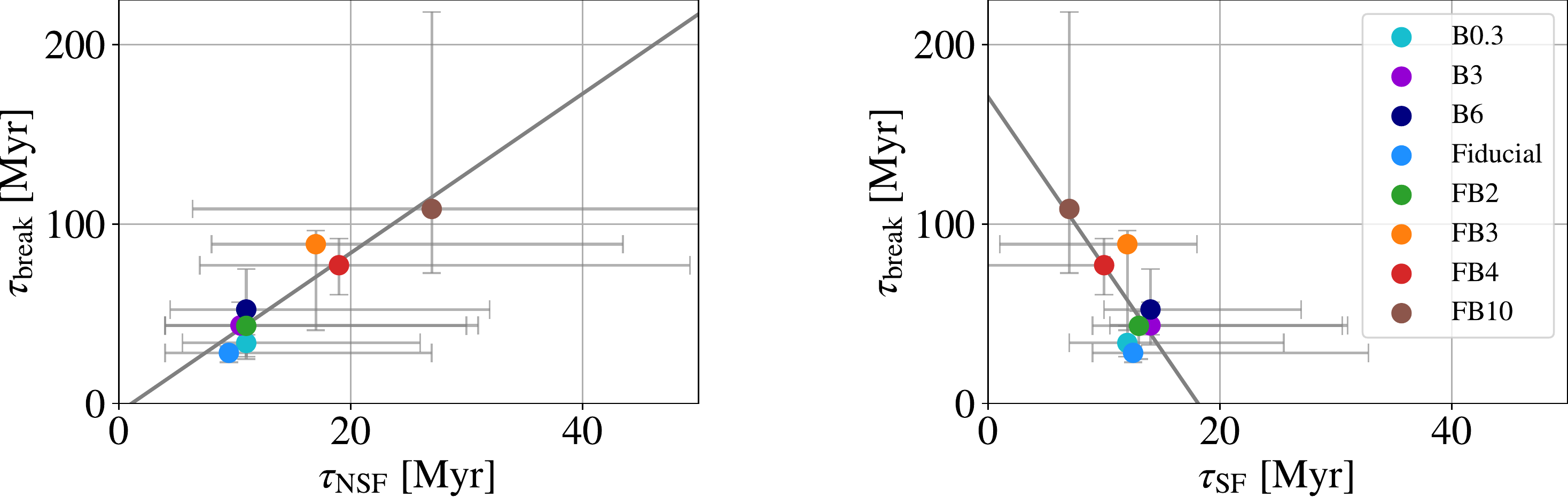}
\caption{
How is star formation variability connected to baryon cycling? 
We plot the relation between the median period of gas in the diffuse ({\it left} panel) and dense ({\it right} panel) ISM phase ($\tau_{\rm NSF}$ and $\tau_{\rm SF}$; presented in Figure~\ref{fig:gas_density_evolution}) and timescale of the break of the PSD ($\tau_{\rm break}$: presented in Figure~\ref{fig:psd-amplitude}) for eight setups.
The error bar indicates 16th and 84th percentiles.
The break time in PSD is correlated with the lifetime of gas in the diffuse ISM phase.
See Section~\ref{subsec:temporal_PSD} for details.}
\label{fig:tb-vs-tnsf}
\end{figure*}

Figure~\ref{fig:tb-vs-tnsf} presents one of the key results of this paper: the correlation between the median break time ($\tau_{\rm break}$) of the temporal PSD function (see Figure~\ref{fig:sPSD}) and the median period of gas in the non-star-forming state ($\tau_{\rm NSF}$) and star-forming state ($\tau_{\rm SF}$; see Figure~\ref{fig:gas_density_evolution}.
The error bars of $\tau_{\rm NSF}$ become larger with the stronger feedback, while $\tau_{\rm SF}$ exhibits reversal tendencies, both of which are due to a wider diversity of gas phases throughout the galaxy, demonstrating the difference in the baryon cycle between different setups.

We find a strong correlation between the break timescale $\tau_{\rm break}$ and the residence time of the gas in the non-star-forming phase ($\tau_{\rm NSF}$) and star-forming phase ($\tau_{\rm SF}$), with the slope of $4.44$ and $-9.43$, respectively.
This shows that the star formation variability---a measure on global scales over the whole galaxy---can be directly related to the baryon cycle internal to galaxies. 
This demonstrates an interesting connection between the global star formation in the galaxies and the temporal evolution of the gas in the ISM.

\section{Discussion}
\label{sec:discussion}

\subsection{Implications for the baryon cycle within galaxies}
\label{subsec:baryon cycle}

\begin{figure*}
\centering
\includegraphics[width=15cm]{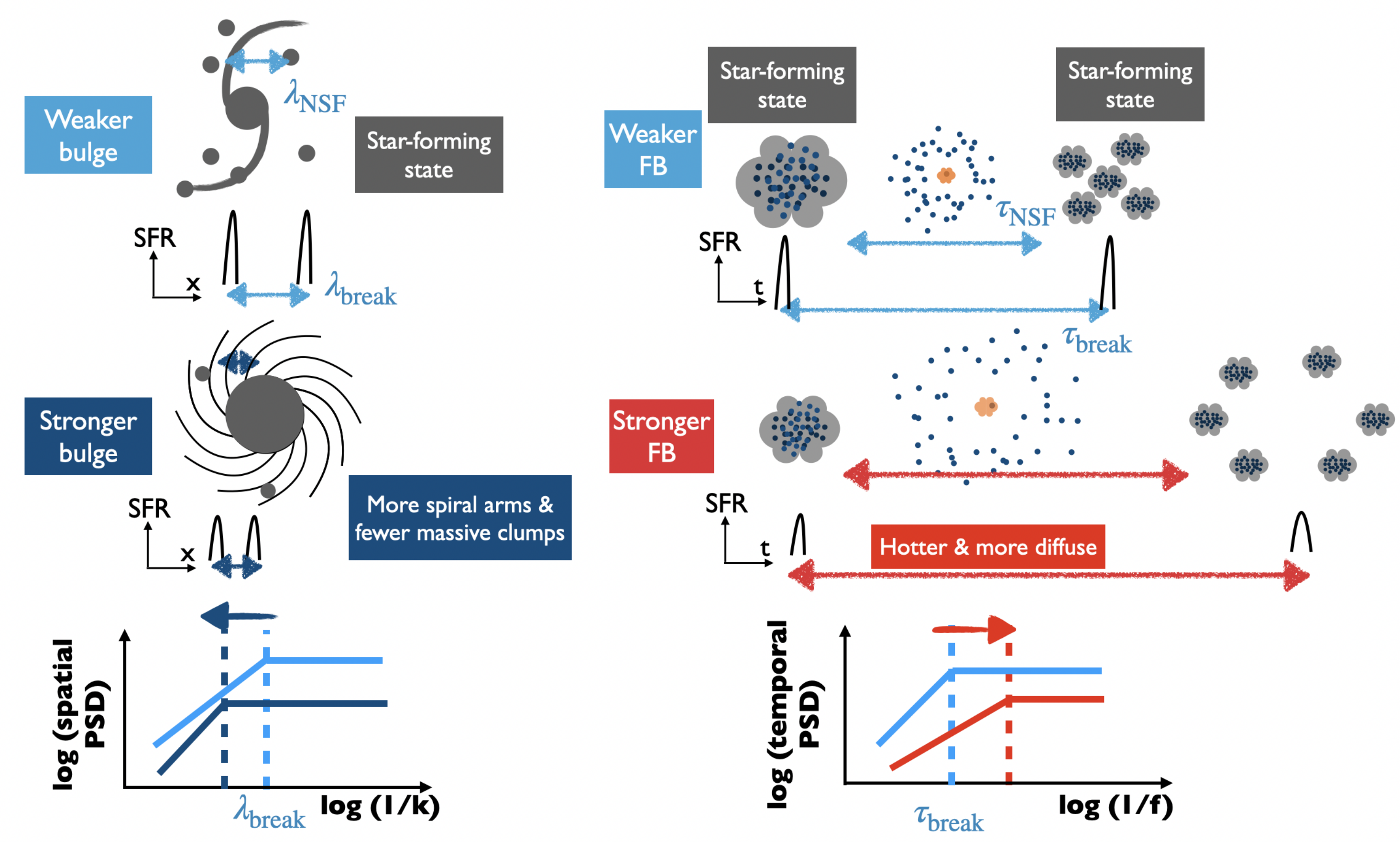}
\caption{
Schematic summary of how bulge mass and feedback influence the gas cycle and spatial and temporal PSDs.
The {\it left} illustrations show the spatial distribution of star-forming gas in galaxies with different bulge mass  and their spatial PSDs, while the {\it right} illustrations show the evolution between non-star-forming and star-forming phase for a specific gas parcel and the temporal PSDs.
The bulge-dominated galaxies form more spiral arms and fewer clumps, resulting in a shorter correlation length of star-forming region and a suppressed SFR.
The ISM gas requires longer time to return to the star-forming state for the strong feedback case, which makes the SFH correlated over longer timescales and is consistent with a larger timescale for the break in the SFH PSD.
See Section~\ref{subsec:baryon cycle} for details.}
\label{fig:cartoon}
\end{figure*}


Since star formation occurs in dense molecular gas, the SFH follows the mass history of dense molecular gas clouds, which is related to the formation and destruction of the GMCs in the galaxy. 
Figure~\ref{fig:cartoon} illustrates the ISM gas distribution and the relationship between the formation timescale of gas clumps and feedback strength.
Following \cite{Semenov2017}, we divide the gas distribution into two states, star-forming and non-star-forming state using a density threshold, $\rho_{\rm SF, thres}$.
In this cartoon, we consider the spatial distribution of star-forming gas, i.e., the gas in spiral arms and clumps, for galaxies with different bulge masses, with the SFRs as a function of the position and the evolution of gas in dense molecular clouds and the SFRs in them as a function of the time for the different strength of feedbacks.

A massive bulge induces strong shear on the ISM and results in a larger number of spiral arms with suppressed density (shown in Section~\ref{subsec:spiral} and ~\ref{subsec:clump}), which leads to the decrease in the correlation length of SFR shown in the spatial PSD.
Dense molecular clouds form stars and $5$ Myr later, the stellar feedback from newly born stars injects thermal energy into the ISM, dispersing the dense gas and rendering it non-star-forming.
For a stronger stellar feedback, gas becomes hotter and more diffuse; consequently, the cooling time increases.
As demonstrated in Section~\ref{subsec:temporal_PSD}, the correlation time of SFR is associated with the residence time in the non-star-forming state, or the formation time of dense gaseous clumps. Therefore, the stronger the feedback, the longer the break timescale of the SFH PSD.

\subsection{Observational implications}

We investigated how the temporal and spatial PSDs of gas and stars varies in galaxies with different stellar thermal feedback strengths and  $\mathrm{B/T}$ (see Figures~\ref{fig:sPSD} and \ref{fig:sfh_psd}). 
We find that the location of the PSD break ($\tau_{\rm break}$) around 10 -- 100 Myr is associated with the lifetime of gas in the diffuse phase, which depends on the stellar feedback strength.
This suggests that observational estimates of the PSD can be used to constrain the strength of stellar feedback and may serve as a probe of the thermodynamical structure of the ISM in galaxies.

\cite{Caplar+Tacchella2019} estimate the break of PSD to be $\tau_{\rm break}=178^{+104}_{-66}$ Myr, using measurements of the main sequence scatter at z $\sim$ 0 and $M_{\star}\sim10^{10}\msun$ from SFR tracers that probe different timescales (i.e., H$_{\alpha}$, UV, IR, and the $u$-band). 
\cite{Caplar+Tacchella2019} and \cite{Tacchella2020} modeled PSD of SFR for various galaxies in different regimes and suggested the break of temporal PSD, $\tau_{\rm break}\sim\,$150 Myr for a Milky Way analogue, which is slightly longer than the longest $\tau_{\rm break}$ among our test.
Note, however, that the entire SFH will be governed by the equilibrium time scale, and $\tau_{\rm break}$ can be longer due to massive clump formation caused by the inflow of gas, which we do not model in our simulations.
Using MaNGA data, \cite{Wang+Lilly2020} reported that the slope of PSD SFH for the local galaxies lies in 1.0 -- 2.0 in the time scale range $5 \,{\rm Myr} - 800\, {\rm Myr}$, which is similar to our results. 
Note however that this result might also be sensitive to presence of the cosmological inflow.

Individual galaxy spectral energy distributions (SEDs) can be used to determine their overall SFH shape \citep{Pacifici+2013, Leja+2017, Iyer+2019}, however, the distribution of spectral features for populations of galaxies contains features that are sensitive to the properties of the temporal PSD \citep{Iyer+2022}. 
Assuming a parametric model for the PSD, it is then possible to infer the timescales and strength of breaks in the PSD using distributions of spectral features including H$\alpha$ luminosity, D$_n(4000)$ break strength, and H$\delta_{\rm EW}$ \citep{Iyer+2022}. 
In practice, this is made challenging due to additional systematics such as poorly constrained metallicity evolution with time, inclination-dependent dust attenuation, IMF variability, and stellar population synthesis assumptions. 
Since we have demonstrated a correlation between spatial and temporal PSD, this study indicates that we can combine information from galaxy SEDs and their spatial PSDs to better constrain the effects of feedback and baryon cycling in galaxies.
A joint spatio-temporal model for star formation variability will provide key constraints on modeling the feedback strengths for the next generation of high-resolution cosmological simulations.

\subsection{Caveats of the analysis}

Since we focus on idealized simulations of isolated galaxies, our analysis provides limited insight into the effect of inflows on the SFH. 
\cite{Iyer2020} investigated the PSDs of cosmological simulations and found that the environment shapes the long-term variability: SFHs of galaxies are coherent with the dark matter accretion histories on long timescales ($\sim5$ Gyr). 
In our study, we focused on the influence of thermal feedback and morphological properties on the galaxies in a controlled setup.

The mass and spatial resolution of our simulations is rather moderate and allows us to probe spatial and temporal PSDs only down to 80 pc and $\sim10$ Myr. 
Our key results and conclusions focus on scales above those limits and therefore are not affected by resolution. We quantify resolution effects on PSD in Appendix~\ref{appsec:resolution}.
Our key results and conclusions focus on scales above those limits and are therefore not affected by resolution. 

We adopt a constant star formation efficiency per free-fall time, $\epsilon_*$, in our star formation prescription (Equation~\ref{eq:SF}). However, an intrinsic variation of $\epsilon_*$ might produce additional spatial and temporal variability of the SFR. Examples of such models where the spatial and temporal variation of $\epsilon_*$ is caused by the dependence on the local turbulent state of gas were explored in \citet{Braun+Schmidt2015,Semenov2016,Kretschmer+Teyssier2020}. 
In the future, it would be interesting to explore how different star formation efficiency models influence the star formation variability. 

Finally, we have only accounted for core-collapse supernovae with the pure thermal feedback model and ignored any other type of stellar feedback or subgrid strategies to alleviate the overcooling problem, such as radiation from young stars \citep{Kim+2013, Kimm+Cen2014}, kinetic feedback \citep{Hopkins+2018}, or stochastic feedback \citep{DallaVecchia2012,Oku2022}.
Implementation of this simple model provides us with a clear view of the impact of thermal energy injection. 
However, this leads to an overcooling of ISM and underestimation of the energy transfer to the ISM via stellar feedback \citep[e.g.,][]{Katz1992,Hu2019}. 
Although our suite of simulations focus mainly on the thermal feedback strength, our ideas and conclusions can be qualitatively generalized since the variation of the feedback strength in our model does produce significant effect on the SFR magnitude and variability. Feedback implementations designed to mitigate overcooling are expected to produce even stronger effects.

\section{Conclusions}
\label{sec:conclusions}

Using isolated galaxy hydrodynamic simulations, we have investigated the connection between the baryon cycle and the star formation variability in the Milky Way-mass galaxies. 
We setup and run 80 simulations that include star formation, stellar feedback and self-consistent cooling and heating. 
Varying the thermal energy of stellar feedback and the $\mathrm{B/T}$ ratio, we quantify how the thermodynamical and morphological properties impact on the baryon cycle and star formation variability.

We measure the amplitude and number of spiral arms and the mass function and stellar fraction of the star-forming clumps as a function of thermal feedback strength and bulge mass (Section~\ref{subsec:spiral} and~\ref{subsec:clump}).
Stronger thermal stellar feedback disperses the dense gas clumps and spiral arms and inhibits the formation of dense structures. 
Meanwhile, a massive bulge exerts a strong shear force; the density amplitude, the number of spiral arms and gas clumps are all significantly affected.

Tracking the time evolution of individual gas parcels, we classify different gas density histories using the {\sc Umap} algorithm and analyze the physical properties of each group (density and spatial movement), how gas moves and oscillates between different phases of the ISM, and forms stars. 
We use our simulations to explain various evolutionary timescales of gas, showing that these timescales are related to the movement through spiral arms and depend on the location in the galaxies.

The spatial and temporal fluctuations of the SFH (as measured by the PSD, see Sections \ref{subsec:spatial_psd} and \ref{subsec:temporal_PSD}) depend significantly on the thermal feedback energy strength and $\mathrm{B/T}$ ratio. 
Strong feedback suppresses fluctuations on all scales in both the temporal and spatial PSD, while the dependence on $\mathrm{B/T}$ emerges on scale of about 0.3 -- 5 kpc for the spatial PSD and on 10 -- 100 Myr for the temporal PSD.
We find that the locations of the break of the PSDs, which correspond to the scale below which the SFH is correlated, shift to larger scales for spatial and temporal PSD with the stronger feedback.
On the other hand, as $\mathrm{B/T}$ increases, the correlation length (spatial PSD) decreases while the correlation time (temporal PSD) remains relatively constant.
We demonstrate that the correlation time of SFH is directly related to the time the gas spends in the diffuse phase (residential time $\tau_{\rm NSF}$), which implies that the SFH PSD is a probe for the baryon cycle within galaxies and the inner dynamical and thermodynamical ISM structure. Hence, the observational measurement of the PSD can provide us a constraint on stellar feedback.


\begin{acknowledgments}
Ji-hoon Kim acknowledges support by Samsung Science and Technology Foundation under Project Number SSTF-BA1802-04, and by the POSCO Science Fellowship of POSCO TJ Park Foundation. His work was also supported by the National Institute of Supercomputing and Network/Korea Institute of Science and Technology Information with supercomputing resources including technical support, grants KSC-2020-CRE-0219 and KSC-2021-CRE-0442.
Support for V.S. was provided by NASA through the NASA Hubble Fellowship grant HST-HF2-51445.001-A awarded by the Space Telescope Science Institute, which is operated by the Association of Universities for Research in Astronomy, Inc., for NASA, under contract NAS5-26555, and by Harvard University through the ITC Fellowship.
\end{acknowledgments}

%



\software{
	{\tt yt-toolkit} \citep{Turk2011},
   {\sc Grackle} \citep{Smith2017},
    {\tt scipy} \citep{scipy},
	{\sc Umap} \citep{McInnes2018}
}



\appendix

\section{Effect of varying the resolution}
\label{appsec:resolution}

\begin{figure*}
\centering
\includegraphics[width=18cm]{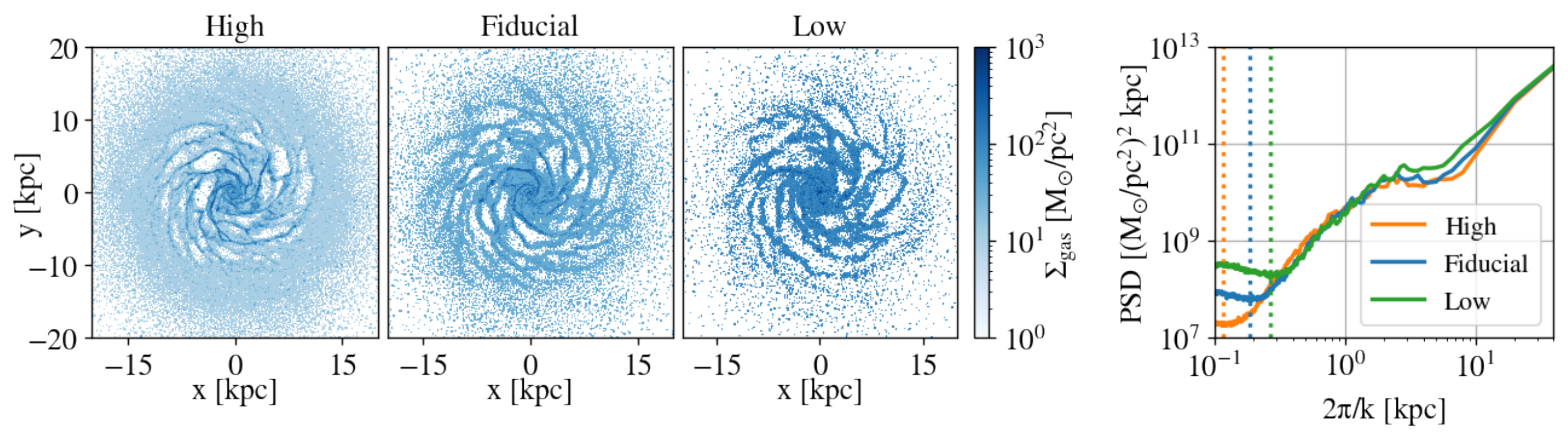}
\caption{
Dependence of the gas spatial distribution on the resolution of the simulation. 
We plot the gas surface density at $t=$ 200 Myr ($\Sigma_{\rm gas}$) of {\tt High} ({\it left most} panel), {\tt Fiducial} ({\it 2nd} panel), and {\tt Low} ({\it 3rd} panel) resolution, and the spatial PSDs of $\Sigma_{\rm gas}$ ({\it right most} panel). 
The vertical dashed lines show the location where the quantization noise exceeds the signals.
Spatial PSD of $\Sigma_{\rm gas}$ on $>0.3$ kpc scales is only weakly sensitive to the resolution. On smaller scales, the PSD flattens, with the transition scale and PSD amplitude both decreasing at higher resolution. See Appendix~\ref{appsec:resolution} for details.
}
\label{fig:resolution-sPSD}
\end{figure*}

\begin{figure*}
\centering
\includegraphics[width=16cm]{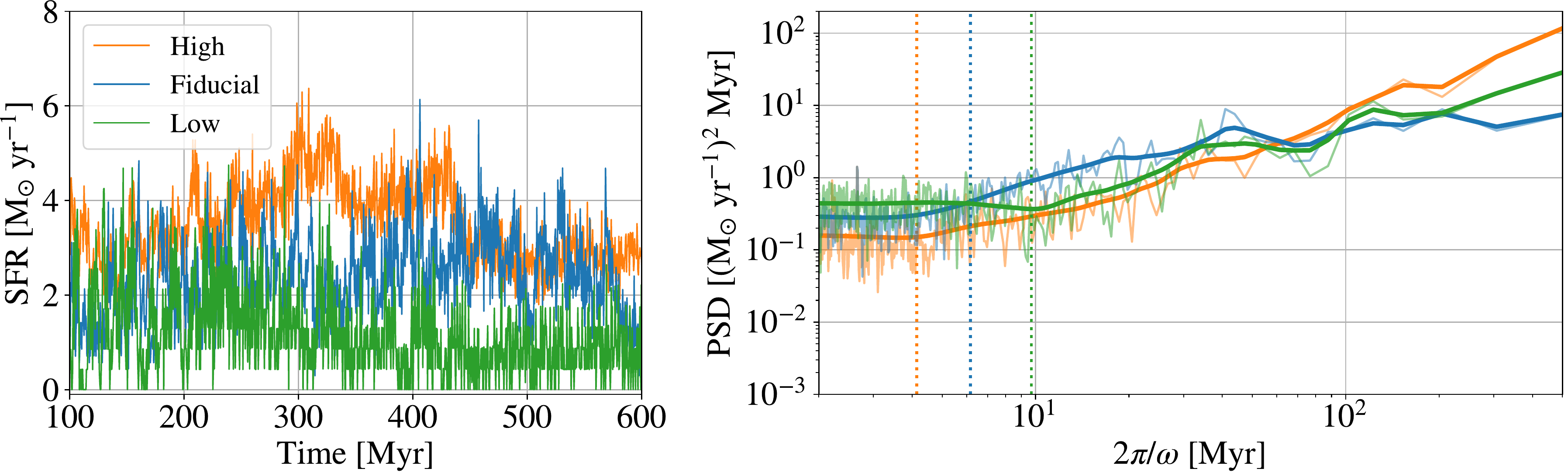}
\caption{
Dependence of the SFHs on the resolution of the simulation.
We plot one individual SFHs of {\tt High}, {\tt Fiducial}, and {\tt Low} runs ({\it left} panel), and the median temporal PSD of 10 runs of a given setup  PSDs of SFR ({\it right} panel).
The vertical dashed lines present the estimated timescale based on Eq.~\ref{eq:t_res}, where the white noise exceeds the signals.
The higher resolution simulation leads to a lower amplitude of the white noise presented in the short fluctuation. 
Therefore, the higher resolution enables us to probe short timescale fluctuations. 
See Appendix~\ref{appsec:resolution} for details.
}
\label{fig:resolution-tPSD}
\end{figure*}

In this section, we present a resolution test to investigate the temporal and spatial PSD dependence on the resolution.
We compare three different simulations: {\tt High}, {\tt Fiducial} and {\tt Low} run.
We employ three times more or less particles for {\tt High} or {\tt Low} runs, compared to {\tt Fiducial} run, so the mass resolution is three times better and worse. 
We adopt the gravitational softening length of $\epsilon_{\rm grav}=$ 56, 80 and 113 pc for {\tt High}, {\tt Fiducial} and {\tt Low} run, respectively. 
We use the same input parameters for the initial condition or star formation with {\tt Fiducial} run.

In Figure~\ref{fig:resolution-sPSD}, we display the gas projections at $t=$ 200 Myr and their spatial PSD with different resolution simulation runs. 
Comparing the projections between three different resolution runs, one can note that the galaxy outskirts are different with the resolution.
{\tt Fiducial} and {\tt low} describe the ISM between the spiral arm has similar density with the out skirt while {\tt High} runs shows more diffuse ISM for the disk near by spiral arm, since the higher resolution simulation allows to describe both denser and diffuse phases in detail.

The PSD functions are reasonably converged on $>0.3$ kpc scales, however, there is a clear difference on smaller scales.
This resolution dependence is due to the quantization noise, as discussed in Section~\ref{subsec:spatial_psd}.
The location where the quantization noise exceeds the signals is shown with the vertical dashed lines: 0.13 ({\tt High}), 0.22 ({\tt Fiducial}), 0.31 ({\tt Low}) kpc.
The ratio between these number is consistent with the change in spatial resolution: 56, 80 and 113 pc.
For small fluctuation scale, the number of particles in a given time bin is less in the lower resolution run than in the higher resolution, resulting in stochastic burstiness in the small scale.
Therefore, the higher resolution simulation reduces the stochastic noise and this allows to look at the small scale signal in PSD. 

Figure~\ref{fig:resolution-tPSD} shows the SFH in different resolution runs and their temporal PSD. 
The overall SFR in the {\tt High} run is higher and {\tt Low} is lower than {\tt Fiducial} run.
As we discussed in the temporal PSD analysis, the stochastic effect of SFH is reduced for the higher resolution. 
Therefore, the higher resolution simulation allows us to probe shorter timescales in PSD.

\cite{Iyer2020} discussed the dependence of the probing limit of the fluctuation time scale on the resolution of the simulations.
Using Figure A2 in \cite{Iyer2020}, we can extrapolate the limit of the fluctuation to the resolution scale of our simulations.
Since our simulations have SFR scale of 1 $\msun$/yr, this gives $\tau_{\rm res}=180~\mathrm{Myr}$ for $m_{\star}\approx10^6~\msun$.
For this work, we estimate the limit of  the fluctuation time scales with,

\begin{equation}
    \tau_{\rm res} = 15\,{\rm Myr} \left(\frac{m_{\star}}{10^6~\msun}\right)^{1/4} \left(\frac{\langle \mathrm{SFR} \rangle}{1~\msun~\mathrm{yr}^{-1}}\right)^{-1/3},
\label{eq:t_res}
\end{equation}
where $m_{\star}$ is the mass of the star particle.

\section{PSD Fitting}
\label{appsec:psd_fitting}

We examine our temporal PSD of SFH in simulation using the Extended Regulator model defined by \cite{Tacchella2020}.
The Regulator model is built on the idea that the SFR variability is driven by the mass of the gas reservoir in the galaxy.
Based on the fundamental mass conservation equation of the gas reservoir, including the source and sink of gas, the Regulator model links together the cosmic inflow and outflow to the inner cycles of the gas in the galaxy.
On top of the Regulator model, the Extended Regulator model accounts for the GMC formation as a source of SFR variability, which is also regulated by the gas reservoir with formation efficiency and lifetime.
The model is defined as:
\begin{equation}
    PSD(f) = PSD_{\rm reg}(f) + PSD_{\rm GMC}(f)
    = \left(\frac{2\sigma^2}{(1+(2\pi\tau_{\rm eq}f)^2)(1+(2\pi\tau_{\rm x}f)^2))}\right)
    + \left(\frac{2\sigma^2_{\rm GMC}}{(1+(2\pi\tau_{\rm L}f)^2))}\right),
\end{equation}
where $\sigma$, $\sigma_{\rm GMC}$ are the normalization amplitudes of the PSD of regulator and GMC model, which are the variance of the signal regarding the each process, $\tau_{\rm eq}$, $\tau_{\rm x}$, and $\tau_{\rm L}$ are the equilibrium timescale, break timescale of regulator model, and GMC life time of GMC model, respectively.
We use Markov Chain Monte Carlo (MCMC) sampler to fit a model to the PSD and find the model these five parameters, $\sigma$, $\tau_{\rm eq}$, $\tau_{\rm x}$, $\sigma_{\rm GMC}$ and $\tau_{\rm L}$ and test other parameters running MCMC on the data to be sure the fitting model evaluating the likelihood of the model.


\bibliography{sample631}{}

\begin{thebibliography}{}
\expandafter\ifx\csname natexlab\endcsname\relax\def\natexlab#1{#1}\fi
\providecommand{\url}[1]{\href{#1}{#1}}
\providecommand{\dodoi}[1]{doi:~\href{http://doi.org/#1}{\nolinkurl{#1}}}
\providecommand{\doeprint}[1]{\href{http://ascl.net/#1}{\nolinkurl{http://ascl.net/#1}}}
\providecommand{\doarXiv}[1]{\href{https://arxiv.org/abs/#1}{\nolinkurl{https://arxiv.org/abs/#1}}}

\bibitem[{{Angl{\'e}s-Alc{\'a}zar} {et~al.}(2014){Angl{\'e}s-Alc{\'a}zar},
  {Dav{\'e}}, {{\"O}zel}, \& {Oppenheimer}}]{Angles-Alcazar+2014}
{Angl{\'e}s-Alc{\'a}zar}, D., {Dav{\'e}}, R., {{\"O}zel}, F., \& {Oppenheimer},
  B.~D. 2014, \apj, 782, 84, \dodoi{10.1088/0004-637X/782/2/84}

\bibitem[{{Angl{\'e}s-Alc{\'a}zar} {et~al.}(2017){Angl{\'e}s-Alc{\'a}zar},
  {Faucher-Gigu{\`e}re}, {Kere{\v{s}}}, {Hopkins}, {Quataert}, \&
  {Murray}}]{Angles-Alcazar+2017}
{Angl{\'e}s-Alc{\'a}zar}, D., {Faucher-Gigu{\`e}re}, C.-A., {Kere{\v{s}}}, D.,
  {et~al.} 2017, \mnras, 470, 4698, \dodoi{10.1093/mnras/stx1517}

\bibitem[{{Bate} \& {Bonnell}(2005)}]{Bate+Bonnell2005}
{Bate}, M.~R., \& {Bonnell}, I.~A. 2005, \mnras, 356, 1201,
  \dodoi{10.1111/j.1365-2966.2004.08593.x}

\bibitem[{{Bouch{\'e}} {et~al.}(2010){Bouch{\'e}}, {Dekel}, {Genzel}, {Genel},
  {Cresci}, {F{\"o}rster Schreiber}, {Shapiro}, {Davies}, \&
  {Tacconi}}]{Bouche2010}
{Bouch{\'e}}, N., {Dekel}, A., {Genzel}, R., {et~al.} 2010, \apj, 718, 1001,
  \dodoi{10.1088/0004-637X/718/2/1001}

\bibitem[{{Braun} \& {Schmidt}(2012)}]{Braun+Schmidt2012}
{Braun}, H., \& {Schmidt}, W. 2012, \mnras, 421, 1838,
  \dodoi{10.1111/j.1365-2966.2011.19889.x}

\bibitem[{{Braun} \& {Schmidt}(2015)}]{Braun+Schmidt2015}
---. 2015, \mnras, 454, 1545, \dodoi{10.1093/mnras/stv1856}

\bibitem[{{Caplar} \& {Tacchella}(2019)}]{Caplar+Tacchella2019}
{Caplar}, N., \& {Tacchella}, S. 2019, \mnras, 487, 3845,
  \dodoi{10.1093/mnras/stz1449}

\bibitem[{{Chabrier}(2003)}]{Chabrier2003}
{Chabrier}, G. 2003, \apjl, 586, L133, \dodoi{10.1086/374879}

\bibitem[{{Christensen} {et~al.}(2016){Christensen}, {Dav{\'e}}, {Governato},
  {Pontzen}, {Brooks}, {Munshi}, {Quinn}, \& {Wadsley}}]{Christensen+2016}
{Christensen}, C.~R., {Dav{\'e}}, R., {Governato}, F., {et~al.} 2016, \apj,
  824, 57, \dodoi{10.3847/0004-637X/824/1/57}

\bibitem[{{Dalla Vecchia} \&
  {Schaye}(2012{\natexlab{a}})}]{DallaVecchia+Schaye2012}
{Dalla Vecchia}, C., \& {Schaye}, J. 2012{\natexlab{a}}, \mnras, 426, 140,
  \dodoi{10.1111/j.1365-2966.2012.21704.x}

\bibitem[{{Dalla Vecchia} \& {Schaye}(2012{\natexlab{b}})}]{DallaVecchia2012}
---. 2012{\natexlab{b}}, \mnras, 426, 140,
  \dodoi{10.1111/j.1365-2966.2012.21704.x}

\bibitem[{{Dav{\'e}} {et~al.}(2012){Dav{\'e}}, {Finlator}, \&
  {Oppenheimer}}]{Dave2012}
{Dav{\'e}}, R., {Finlator}, K., \& {Oppenheimer}, B.~D. 2012, \mnras, 421, 98,
  \dodoi{10.1111/j.1365-2966.2011.20148.x}

\bibitem[{{Dekel} \& {Mandelker}(2014)}]{Dekel2014}
{Dekel}, A., \& {Mandelker}, N. 2014, \mnras, 444, 2071,
  \dodoi{10.1093/mnras/stu1427}

\bibitem[{{Dobbs} {et~al.}(2022){Dobbs}, {Bending}, {Pettitt}, \&
  {Bate}}]{Dobbs+2022}
{Dobbs}, C.~L., {Bending}, T.~J.~R., {Pettitt}, A.~R., \& {Bate}, M.~R. 2022,
  \mnras, 509, 954, \dodoi{10.1093/mnras/stab3036}

\bibitem[{{Eisenstein} \& {Hut}(1998)}]{Eisenstein1998}
{Eisenstein}, D.~J., \& {Hut}, P. 1998, \apj, 498, 137, \dodoi{10.1086/305535}

\bibitem[{{Elmegreen}(2011)}]{Elmegreen2011}
{Elmegreen}, B.~G. 2011, in EAS Publications Series, Vol.~51, EAS Publications
  Series, ed. C.~{Charbonnel} \& T.~{Montmerle}, 19--30,
  \dodoi{10.1051/eas/1151002}

\bibitem[{{Emerick} {et~al.}(2018){Emerick}, {Bryan}, \& {Mac
  Low}}]{Emerick+2018}
{Emerick}, A., {Bryan}, G.~L., \& {Mac Low}, M.-M. 2018, \apjl, 865, L22,
  \dodoi{10.3847/2041-8213/aae315}

\bibitem[{{Faucher-Gigu{\`e}re} {et~al.}(2011){Faucher-Gigu{\`e}re},
  {Kere{\v{s}}}, \& {Ma}}]{Faucher-Giguere+2011}
{Faucher-Gigu{\`e}re}, C.-A., {Kere{\v{s}}}, D., \& {Ma}, C.-P. 2011, \mnras,
  417, 2982, \dodoi{10.1111/j.1365-2966.2011.19457.x}

\bibitem[{{Ferland} {et~al.}(2013){Ferland}, {Porter}, {van Hoof}, {Williams},
  {Abel}, {Lykins}, {Shaw}, {Henney}, \& {Stancil}}]{Ferland2013}
{Ferland}, G.~J., {Porter}, R.~L., {van Hoof}, P.~A.~M., {et~al.} 2013, \rmxaa,
  49, 137.
\newblock \doarXiv{1302.4485}

\bibitem[{{Fielding} {et~al.}(2017){Fielding}, {Quataert}, {McCourt}, \&
  {Thompson}}]{Fielding+2017}
{Fielding}, D., {Quataert}, E., {McCourt}, M., \& {Thompson}, T.~A. 2017,
  \mnras, 466, 3810, \dodoi{10.1093/mnras/stw3326}

\bibitem[{{F{\"o}rster Schreiber} {et~al.}(2014){F{\"o}rster Schreiber},
  {Genzel}, {Newman}, {Kurk}, {Lutz}, {Tacconi}, {Wuyts}, {Bandara}, {Burkert},
  {Buschkamp}, {Carollo}, {Cresci}, {Daddi}, {Davies}, {Eisenhauer}, {Hicks},
  {Lang}, {Lilly}, {Mainieri}, {Mancini}, {Naab}, {Peng}, {Renzini}, {Rosario},
  {Shapiro Griffin}, {Shapley}, {Sternberg}, {Tacchella}, {Vergani},
  {Wisnioski}, {Wuyts}, \& {Zamorani}}]{Forster-Schreiber+2014}
{F{\"o}rster Schreiber}, N.~M., {Genzel}, R., {Newman}, S.~F., {et~al.} 2014,
  \apj, 787, 38, \dodoi{10.1088/0004-637X/787/1/38}

\bibitem[{{F{\"o}rster Schreiber} {et~al.}(2019){F{\"o}rster Schreiber},
  {{\"U}bler}, {Davies}, {Genzel}, {Wisnioski}, {Belli}, {Shimizu}, {Lutz},
  {Fossati}, {Herrera-Camus}, {Mendel}, {Tacconi}, {Wilman}, {Beifiori},
  {Brammer}, {Burkert}, {Carollo}, {Davies}, {Eisenhauer}, {Fabricius},
  {Lilly}, {Momcheva}, {Naab}, {Nelson}, {Price}, {Renzini}, {Saglia},
  {Sternberg}, {van Dokkum}, \& {Wuyts}}]{Forster-Schreiber+2019}
{F{\"o}rster Schreiber}, N.~M., {{\"U}bler}, H., {Davies}, R.~L., {et~al.}
  2019, \apj, 875, 21, \dodoi{10.3847/1538-4357/ab0ca2}

\bibitem[{{Fraternali}(2017)}]{Fraternali2017}
{Fraternali}, F. 2017, in Astrophysics and Space Science Library, Vol. 430, Gas
  Accretion onto Galaxies, ed. A.~{Fox} \& R.~{Dav{\'e}}, 323,
  \dodoi{10.1007/978-3-319-52512-9\_14}

\bibitem[{{Gensior} {et~al.}(2020){Gensior}, {Kruijssen}, \&
  {Keller}}]{Gensior2020}
{Gensior}, J., {Kruijssen}, J.~M.~D., \& {Keller}, B.~W. 2020, \mnras, 495,
  199, \dodoi{10.1093/mnras/staa1184}

\bibitem[{{Goh} \& {Barab{\'a}si}(2008)}]{Goh+2008}
{Goh}, K.~I., \& {Barab{\'a}si}, A.~L. 2008, EPL (Europhysics Letters), 81,
  48002, \dodoi{10.1209/0295-5075/81/48002}

\bibitem[{{Guszejnov} {et~al.}(2021){Guszejnov}, {Grudi{\'c}}, {Hopkins},
  {Offner}, \& {Faucher-Gigu{\`e}re}}]{Guszejnov+2021}
{Guszejnov}, D., {Grudi{\'c}}, M.~Y., {Hopkins}, P.~F., {Offner}, S. S.~R., \&
  {Faucher-Gigu{\`e}re}, C.-A. 2021, \mnras, 502, 3646,
  \dodoi{10.1093/mnras/stab278}

\bibitem[{{Haardt} \& {Madau}(2012)}]{HaardtMadau12}
{Haardt}, F., \& {Madau}, P. 2012, \apj, 746, 125,
  \dodoi{10.1088/0004-637X/746/2/125}

\bibitem[{{Heckman} \& {Thompson}(2017)}]{Heckman+Thompson2017}
{Heckman}, T.~M., \& {Thompson}, T.~A. 2017, arXiv e-prints, arXiv:1701.09062.
\newblock \doarXiv{1701.09062}

\bibitem[{{Hennebelle} \& {Chabrier}(2008)}]{Hennebelle+Chabrier2008}
{Hennebelle}, P., \& {Chabrier}, G. 2008, \apj, 684, 395,
  \dodoi{10.1086/589916}

\bibitem[{{Hernquist}(1990)}]{Hernquist1990}
{Hernquist}, L. 1990, \apj, 356, 359, \dodoi{10.1086/168845}

\bibitem[{{Hernquist} \& {Katz}(1989)}]{Hernquist1989}
{Hernquist}, L., \& {Katz}, N. 1989, \apjs, 70, 419, \dodoi{10.1086/191344}

\bibitem[{{Hopkins}(2013)}]{Hopkins2013}
{Hopkins}, P.~F. 2013, \mnras, 428, 2840, \dodoi{10.1093/mnras/sts210}

\bibitem[{{Hopkins}(2015)}]{Hopkins2015}
---. 2015, \mnras, 450, 53, \dodoi{10.1093/mnras/stv195}

\bibitem[{{Hopkins} {et~al.}(2012){Hopkins}, {Quataert}, \&
  {Murray}}]{Hopkins+2012}
{Hopkins}, P.~F., {Quataert}, E., \& {Murray}, N. 2012, \mnras, 421, 3522,
  \dodoi{10.1111/j.1365-2966.2012.20593.x}

\bibitem[{{Hopkins} {et~al.}(2018){Hopkins}, {Wetzel}, {Kere{\v{s}}},
  {Faucher-Gigu{\`e}re}, {Quataert}, {Boylan-Kolchin}, {Murray}, {Hayward}, \&
  {El-Badry}}]{Hopkins+2018}
{Hopkins}, P.~F., {Wetzel}, A., {Kere{\v{s}}}, D., {et~al.} 2018, \mnras, 477,
  1578, \dodoi{10.1093/mnras/sty674}

\bibitem[{{Hu}(2019)}]{Hu2019}
{Hu}, C.-Y. 2019, \mnras, 483, 3363, \dodoi{10.1093/mnras/sty3252}

\bibitem[{{Hu} {et~al.}(2017){Hu}, {Naab}, {Glover}, {Walch}, \&
  {Clark}}]{Hu+2017}
{Hu}, C.-Y., {Naab}, T., {Glover}, S. C.~O., {Walch}, S., \& {Clark}, P.~C.
  2017, \mnras, 471, 2151, \dodoi{10.1093/mnras/stx1773}

\bibitem[{{Iyer} {et~al.}(2019){Iyer}, {Gawiser}, {Faber}, {Ferguson},
  {Kartaltepe}, {Koekemoer}, {Pacifici}, \& {Somerville}}]{Iyer+2019}
{Iyer}, K.~G., {Gawiser}, E., {Faber}, S.~M., {et~al.} 2019, \apj, 879, 116,
  \dodoi{10.3847/1538-4357/ab2052}

\bibitem[{{Iyer} {et~al.}(2022){Iyer}, {Speagle}, {Caplar}, {Forbes},
  {Gawiser}, {Leja}, \& {Tacchella}}]{Iyer+2022}
{Iyer}, K.~G., {Speagle}, J.~S., {Caplar}, N., {et~al.} 2022, arXiv e-prints,
  arXiv:2208.05938.
\newblock \doarXiv{2208.05938}

\bibitem[{{Iyer} {et~al.}(2020){Iyer}, {Tacchella}, {Genel}, {Hayward},
  {Hernquist}, {Brooks}, {Caplar}, {Dav{\'e}}, {Diemer}, {Forbes}, {Gawiser},
  {Somerville}, \& {Starkenburg}}]{Iyer2020}
{Iyer}, K.~G., {Tacchella}, S., {Genel}, S., {et~al.} 2020, \mnras, 498, 430,
  \dodoi{10.1093/mnras/staa2150}

\bibitem[{{Katz}(1992)}]{Katz1992}
{Katz}, N. 1992, \apj, 391, 502, \dodoi{10.1086/171366}

\bibitem[{{Keller} {et~al.}(2019){Keller}, {Wadsley}, {Wang}, \&
  {Kruijssen}}]{Keller+2019}
{Keller}, B.~W., {Wadsley}, J.~W., {Wang}, L., \& {Kruijssen}, J.~M.~D. 2019,
  \mnras, 482, 2244, \dodoi{10.1093/mnras/sty2859}

\bibitem[{{Kere{\v{s}}} {et~al.}(2005){Kere{\v{s}}}, {Katz}, {Weinberg}, \&
  {Dav{\'e}}}]{Keres+2005}
{Kere{\v{s}}}, D., {Katz}, N., {Weinberg}, D.~H., \& {Dav{\'e}}, R. 2005,
  \mnras, 363, 2, \dodoi{10.1111/j.1365-2966.2005.09451.x}

\bibitem[{{Kim} \& {Ostriker}(2015)}]{Kim+Ostriker2015}
{Kim}, C.-G., \& {Ostriker}, E.~C. 2015, \apj, 802, 99,
  \dodoi{10.1088/0004-637X/802/2/99}

\bibitem[{{Kim} {et~al.}(2013){Kim}, {Krumholz}, {Wise}, {Turk}, {Goldbaum}, \&
  {Abel}}]{Kim+2013}
{Kim}, J.-h., {Krumholz}, M.~R., {Wise}, J.~H., {et~al.} 2013, \apj, 779, 8,
  \dodoi{10.1088/0004-637X/779/1/8}

\bibitem[{{Kim} {et~al.}(2016){Kim}, {Agertz}, {Teyssier}, {Butler},
  {Ceverino}, {Choi}, {Feldmann}, {Keller}, {Lupi}, {Quinn}, {Revaz},
  {Wallace}, {Gnedin}, {Leitner}, {Shen}, {Smith}, {Thompson}, {Turk}, {Abel},
  {Arraki}, {Benincasa}, {Chakrabarti}, {DeGraf}, {Dekel}, {Goldbaum},
  {Hopkins}, {Hummels}, {Klypin}, {Li}, {Madau}, {Mandelker}, {Mayer},
  {Nagamine}, {Nickerson}, {O'Shea}, {Primack}, {Roca-F{\`a}brega}, {Semenov},
  {Shimizu}, {Simpson}, {Todoroki}, {Wadsley}, {Wise}, \& {AGORA
  Collaboration}}]{Kim+2016}
{Kim}, J.-h., {Agertz}, O., {Teyssier}, R., {et~al.} 2016, \apj, 833, 202,
  \dodoi{10.3847/1538-4357/833/2/202}

\bibitem[{{Kim} {et~al.}(2003){Kim}, {Ostriker}, \& {Stone}}]{Kim+2003}
{Kim}, W.-T., {Ostriker}, E.~C., \& {Stone}, J.~M. 2003, \apj, 599, 1157,
  \dodoi{10.1086/379367}

\bibitem[{{Kimm} \& {Cen}(2014)}]{Kimm+Cen2014}
{Kimm}, T., \& {Cen}, R. 2014, \apj, 788, 121,
  \dodoi{10.1088/0004-637X/788/2/121}

\bibitem[{{Kretschmer} \& {Teyssier}(2020)}]{Kretschmer+Teyssier2020}
{Kretschmer}, M., \& {Teyssier}, R. 2020, \mnras, 492, 1385,
  \dodoi{10.1093/mnras/stz3495}

\bibitem[{{Leja} {et~al.}(2017){Leja}, {Johnson}, {Conroy}, {van Dokkum}, \&
  {Byler}}]{Leja+2017}
{Leja}, J., {Johnson}, B.~D., {Conroy}, C., {van Dokkum}, P.~G., \& {Byler}, N.
  2017, \apj, 837, 170, \dodoi{10.3847/1538-4357/aa5ffe}

\bibitem[{{Li} \& {Tonnesen}(2020)}]{Li+Tonnesen2020}
{Li}, M., \& {Tonnesen}, S. 2020, \apj, 898, 148,
  \dodoi{10.3847/1538-4357/ab9f9f}

\bibitem[{{Lilly} {et~al.}(2013){Lilly}, {Carollo}, {Pipino}, {Renzini}, \&
  {Peng}}]{Lilly2013}
{Lilly}, S.~J., {Carollo}, C.~M., {Pipino}, A., {Renzini}, A., \& {Peng}, Y.
  2013, \apj, 772, 119, \dodoi{10.1088/0004-637X/772/2/119}

\bibitem[{{Martig} {et~al.}(2009){Martig}, {Bournaud}, {Teyssier}, \&
  {Dekel}}]{Martig+2009}
{Martig}, M., {Bournaud}, F., {Teyssier}, R., \& {Dekel}, A. 2009, \apj, 707,
  250, \dodoi{10.1088/0004-637X/707/1/250}

\bibitem[{{McInnes} {et~al.}(2018){McInnes}, {Healy}, \&
  {Melville}}]{McInnes2018}
{McInnes}, L., {Healy}, J., \& {Melville}, J. 2018, arXiv e-prints,
  arXiv:1802.03426.
\newblock \doarXiv{1802.03426}

\bibitem[{{McKee} \& {Ostriker}(2007)}]{McKee+Ostriker2007}
{McKee}, C.~F., \& {Ostriker}, E.~C. 2007, \araa, 45, 565,
  \dodoi{10.1146/annurev.astro.45.051806.110602}

\bibitem[{{Navarro} {et~al.}(1997){Navarro}, {Frenk}, \& {White}}]{Navarro1997}
{Navarro}, J.~F., {Frenk}, C.~S., \& {White}, S. D.~M. 1997, \apj, 490, 493,
  \dodoi{10.1086/304888}

\bibitem[{{Oku} {et~al.}(2022){Oku}, {Tomida}, {Nagamine}, {Shimizu}, \&
  {Cen}}]{Oku2022}
{Oku}, Y., {Tomida}, K., {Nagamine}, K., {Shimizu}, I., \& {Cen}, R. 2022,
  \apjs, 262, 9, \dodoi{10.3847/1538-4365/ac77ff}

\bibitem[{{Oppenheimer} \& {Dav{\'e}}(2008)}]{Oppenheimer+Dave2008}
{Oppenheimer}, B.~D., \& {Dav{\'e}}, R. 2008, \mnras, 387, 577,
  \dodoi{10.1111/j.1365-2966.2008.13280.x}

\bibitem[{{Oppenheimer} {et~al.}(2010){Oppenheimer}, {Dav{\'e}}, {Kere{\v{s}}},
  {Fardal}, {Katz}, {Kollmeier}, \& {Weinberg}}]{Oppenheimer+2010}
{Oppenheimer}, B.~D., {Dav{\'e}}, R., {Kere{\v{s}}}, D., {et~al.} 2010, \mnras,
  406, 2325, \dodoi{10.1111/j.1365-2966.2010.16872.x}

\bibitem[{{Pacifici} {et~al.}(2013){Pacifici}, {Kassin}, {Weiner}, {Charlot},
  \& {Gardner}}]{Pacifici+2013}
{Pacifici}, C., {Kassin}, S.~A., {Weiner}, B., {Charlot}, S., \& {Gardner},
  J.~P. 2013, \apjl, 762, L15, \dodoi{10.1088/2041-8205/762/1/L15}

\bibitem[{{Padoan} \& {Nordlund}(2002)}]{Padoan+Nordlund2002}
{Padoan}, P., \& {Nordlund}, {\r{A}}. 2002, \apj, 576, 870,
  \dodoi{10.1086/341790}

\bibitem[{{Perret}(2016)}]{Perret2016}
{Perret}, V. 2016, {DICE: Disk Initial Conditions Environment}, Astrophysics
  Source Code Library, record ascl:1607.002.
\newblock \doeprint{1607.002}

\bibitem[{{Putman} {et~al.}(2012){Putman}, {Peek}, \& {Joung}}]{Putman2012}
{Putman}, M.~E., {Peek}, J.~E.~G., \& {Joung}, M.~R. 2012, \araa, 50, 491,
  \dodoi{10.1146/annurev-astro-081811-125612}

\bibitem[{{Semenov} {et~al.}(2016){Semenov}, {Kravtsov}, \&
  {Gnedin}}]{Semenov2016}
{Semenov}, V.~A., {Kravtsov}, A.~V., \& {Gnedin}, N.~Y. 2016, \apj, 826, 200,
  \dodoi{10.3847/0004-637X/826/2/200}

\bibitem[{{Semenov} {et~al.}(2017){Semenov}, {Kravtsov}, \&
  {Gnedin}}]{Semenov2017}
---. 2017, \apj, 845, 133, \dodoi{10.3847/1538-4357/aa8096}

\bibitem[{{Semenov} {et~al.}(2018){Semenov}, {Kravtsov}, \&
  {Gnedin}}]{Semenov2018}
---. 2018, \apj, 861, 4, \dodoi{10.3847/1538-4357/aac6eb}

\bibitem[{{Shin} {et~al.}(2021){Shin}, {Kim}, \& {Oh}}]{Shin2021}
{Shin}, E.-J., {Kim}, J.-H., \& {Oh}, B.~K. 2021, \apj, 917, 12,
  \dodoi{10.3847/1538-4357/abffd0}

\bibitem[{{Smith} {et~al.}(2017){Smith}, {Bryan}, {Glover}, {Goldbaum}, {Turk},
  {Regan}, {Wise}, {Schive}, {Abel}, {Emerick}, {O'Shea}, {Anninos}, {Hummels},
  \& {Khochfar}}]{Smith2017}
{Smith}, B.~D., {Bryan}, G.~L., {Glover}, S.~C.~O., {et~al.} 2017, \mnras, 466,
  2217, \dodoi{10.1093/mnras/stw3291}

\bibitem[{{Springel} \& {Hernquist}(2003)}]{Springel+Hernquist2003}
{Springel}, V., \& {Hernquist}, L. 2003, \mnras, 339, 289,
  \dodoi{10.1046/j.1365-8711.2003.06206.x}

\bibitem[{{Stinson} {et~al.}(2013){Stinson}, {Brook}, {Macci{\`o}}, {Wadsley},
  {Quinn}, \& {Couchman}}]{Stinson+2013}
{Stinson}, G.~S., {Brook}, C., {Macci{\`o}}, A.~V., {et~al.} 2013, \mnras, 428,
  129, \dodoi{10.1093/mnras/sts028}

\bibitem[{{Tacchella} {et~al.}(2016){Tacchella}, {Dekel}, {Carollo},
  {Ceverino}, {DeGraf}, {Lapiner}, {Mandelker}, \& {Primack
  Joel}}]{Tacchella2016}
{Tacchella}, S., {Dekel}, A., {Carollo}, C.~M., {et~al.} 2016, \mnras, 457,
  2790, \dodoi{10.1093/mnras/stw131}

\bibitem[{{Tacchella} {et~al.}(2020){Tacchella}, {Forbes}, \&
  {Caplar}}]{Tacchella2020}
{Tacchella}, S., {Forbes}, J.~C., \& {Caplar}, N. 2020, \mnras, 497, 698,
  \dodoi{10.1093/mnras/staa1838}

\bibitem[{{Truelove} {et~al.}(1997){Truelove}, {Klein}, {McKee}, {Holliman},
  {Howell}, \& {Greenough}}]{Truelove1997}
{Truelove}, J.~K., {Klein}, R.~I., {McKee}, C.~F., {et~al.} 1997, \apjl, 489,
  L179, \dodoi{10.1086/310975}

\bibitem[{{Turk} \& {Smith}(2011)}]{Turk2011}
{Turk}, M.~J., \& {Smith}, B.~D. 2011, arXiv e-prints, arXiv:1112.4482.
\newblock \doarXiv{1112.4482}

\bibitem[{{Veilleux} {et~al.}(2005){Veilleux}, {Cecil}, \&
  {Bland-Hawthorn}}]{Veilleux+2005}
{Veilleux}, S., {Cecil}, G., \& {Bland-Hawthorn}, J. 2005, \araa, 43, 769,
  \dodoi{10.1146/annurev.astro.43.072103.150610}

\bibitem[{Virtanen {et~al.}(2020)Virtanen, Gommers, Oliphant, Haberland, Reddy,
  Cournapeau, Burovski, Peterson, Weckesser, Bright, van~der Walt, Brett,
  Wilson, Millman, Mayorov, Nelson, Jones, Kern, Larson, Carey, Polat, Feng,
  Moore, VanderPlas, Laxalde, Perktold, Cimrman, Henriksen, Quintero, Harris,
  Archibald, Ribeiro, Pedregosa, van Mulbregt, Vijaykumar, Bardelli, Rothberg,
  Hilboll, Kloeckner, Scopatz, Lee, Rokem, Woods, Fulton, Masson,
  H{\"{a}}ggstr{\"{o}}m, Fitzgerald, Nicholson, Hagen, Pasechnik, Olivetti,
  Martin, Wieser, Silva, Lenders, Wilhelm, Young, Price, Ingold, Allen, Lee,
  Audren, Probst, Dietrich, Silterra, Webber, Slavi{\v{c}}, Nothman, Buchner,
  Kulick, Sch{\"{o}}nberger, {de Miranda Cardoso}, Reimer, Harrington,
  Rodr{\'{i}}guez, Nunez-Iglesias, Kuczynski, Tritz, Thoma, Newville,
  K{\"{u}}mmerer, Bolingbroke, Tartre, Pak, Smith, Nowaczyk, Shebanov, Pavlyk,
  Brodtkorb, Lee, McGibbon, Feldbauer, Lewis, Tygier, Sievert, Vigna, Peterson,
  More, Pudlik, Oshima, Pingel, Robitaille, Spura, Jones, Cera, Leslie, Zito,
  Krauss, Upadhyay, Halchenko, \& V{\'{a}}zquez-Baeza}]{scipy}
Virtanen, P., Gommers, R., Oliphant, T.~E., {et~al.} 2020, Nature Methods, 17,
  261, \dodoi{10.1038/s41592-019-0686-2}

\bibitem[{{Wang} \& {Lilly}(2020)}]{Wang+Lilly2020}
{Wang}, E., \& {Lilly}, S.~J. 2020, \apj, 895, 25,
  \dodoi{10.3847/1538-4357/ab8b5e}

\bibitem[{{Yepes} {et~al.}(1997){Yepes}, {Kates}, {Khokhlov}, \&
  {Klypin}}]{Yepes+1997}
{Yepes}, G., {Kates}, R., {Khokhlov}, A., \& {Klypin}, A. 1997, \mnras, 284,
  235, \dodoi{10.1093/mnras/284.1.235}

\bibitem[{{Yu} {et~al.}(2018){Yu}, {Ho}, {Barth}, \& {Li}}]{Yu2018}
{Yu}, S.-Y., {Ho}, L.~C., {Barth}, A.~J., \& {Li}, Z.-Y. 2018, \apj, 862, 13,
  \dodoi{10.3847/1538-4357/aacb25}

\end{thebibliography}
\bibliographystyle{aasjournal}



\end{document}